\documentclass[fleqn,usenatbib]{mnras}

\usepackage[T1]{fontenc}

\DeclareRobustCommand{\VAN}[3]{#2}
\let\VANthebibliography\thebibliography
\def\thebibliography{\DeclareRobustCommand{\VAN}[3]{##3}\VANthebibliography}

\usepackage{graphicx}	% Including figure files
\usepackage{amsmath}	% Advanced maths commands
\usepackage{amssymb}	% Extra maths symbols
\usepackage{subcaption}
\usepackage[dvipsnames]{xcolor}
\usepackage{xspace}
\setcitestyle{notesep={ }}
\usepackage{comment}

\newcommand{\mufasa}{{\sc Mufasa}\xspace}
\newcommand{\simba}{{\sc Simba}\xspace}

\newcommand{\lya}{Ly$\alpha$\xspace}
\newcommand{\HI}{\ion{H}{i}\xspace}
\newcommand{\MgII}{\ion{Mg}{ii}\xspace}
\newcommand{\CII}{\ion{C}{ii}\xspace}
\newcommand{\SiIII}{\ion{Si}{iii}\xspace}
\newcommand{\CIV}{\ion{C}{iv}\xspace}
\newcommand{\OVI}{\ion{O}{vi}\xspace}

\newcommand{\hkpc}{h^{-1}{\rm kpc}}

\newcommand{\hmpc}{h^{-1}{\rm Mpc}}
\newcommand{\mpc}{\rm Mpc}

\newcommand{\kms}{\;{\rm km}\,{\rm s}^{-1}}

\newcommand{\msolar}{\;{\rm M}_{\odot}}

\newcommand{\gizmo}{{\sc Gizmo}\xspace}
\newcommand{\caesar}{{\sc Caesar}\xspace}

\newcommand{\fedd}{f_{\rm Edd}}

\newcommand{\mstar}{\mbox{$M_\star$}\xspace}

\newcommand{\pygad}{{\sc Pygad}\xspace}
\newcommand{\music}{{\sc Music}\xspace}
\newcommand{\cloudy}{{\sc Cloudy}\xspace}
\newcommand{\grackle}{{\sc Grackle}\xspace}

\newcommand{\lstar}{L$^\star$\xspace}

\title[CGM absorbers in Simba]{The Physical Nature of Circumgalactic Medium Absorbers in \simba}

\author[S. Appleby et al.]{
Sarah Appleby$^{1}$\thanks{E-mail: sarahappleby20@gmail.com},
Romeel Dav\'e$^{1,2,3}$,
Daniele Sorini$^{1,4,5}$,
Weiguang Cui$^{1,6}$,
Jacob Christiansen$^{1}$
\\
\\$^{1}$ SUPA\thanks{Scottish Universities Physics Alliance}, Institute for Astronomy, University of Edinburgh, Royal Observatory, Edinburgh EH9 3HJ, UK
\\$^2$ University of the Western Cape, Bellville, Cape Town 7535, South Africa
\\$^3$ South African Astronomical Observatories, Observatory, Cape Town 7925, South Africa
\\$^4$D\'epartement de Physique Th\'eorique, Universit\'e de Gen\`eve, 24 quai Ernest Ansermet, 1211 Gen\`eve 4, Switzerland
\\$^5$Institute for Computational Cosmology, Department of Physics, Durham University, South Road, Durham, DH1 3LE, UK
\\$^6$ Departamento de Física Teórica and CIAFF, Modulo 8 Universidad Autónoma de Madrid, 28049 Madrid, Spain.
}

\date{Accepted 2023 January 4. Received 2023 January 4; in original form 2022 July 8}

\pubyear{2023}

\begin{document}
\label{firstpage}
\pagerange{\pageref{firstpage}--\pageref{lastpage}}
\maketitle

\begin{abstract}
% Context, Aims, Methods, Results, and Conclusions.
We study the nature of the low-redshift CGM in the \simba\ cosmological simulations as traced by ultraviolet absorption lines around galaxies in bins of stellar mass ($\mstar>10^{10}M_\odot$) for star-forming, green valley and quenched galaxies at impact parameters $r_\perp\leq 1.25r_{200}$. We generate synthetic spectra for \HI, \MgII, \CII, \SiIII, \CIV, and \OVI, fit Voigt profiles to obtain line properties, and estimate the density, temperature, and metallicity of the absorbing gas.
We find that CGM absorbers are most abundant around star forming galaxies with $\mstar < 10^{11}\msolar$, while the abundance of green valley galaxies show similar behaviour to those of quenched galaxies, suggesting that the CGM ``quenches" before star formation ceases. 
\HI\ absorbing gas exists across a broad range of cosmic phases (condensed gas, diffuse gas, hot halo gas and Warm-Hot Intergalactic Medium), while essentially all low-ionisation metal absorption arises from condensed gas. \OVI absorbers are split between hot halo gas and the WHIM. The fraction of collisionally ionised CGM absorbers is $\sim 25-55\%$ for \CIV\ and $\sim 80-95\%$ for \OVI, depending on stellar mass and impact parameter. In general, the highest column density absorption features for each ion arise from dense gas.
Satellite gas, defined as that within $10r_{1/2,\star},$ contributes $\sim 3\%$ of overall \HI\ absorption but $\sim 30\%$ of \MgII\ absorption, with the fraction from satellites decreasing with increasing ion excitation energy.
\end{abstract}


%%%%%%%%%%%%%%%%%%%%%%%%%%%%%%%%%%%%%%%%%%%%%%%%%%

%%%%%%%%%%%%%%%%% BODY OF PAPER %%%%%%%%%%%%%%%%%%

\section{Introduction}\label{sec:intro}

The circumgalactic medium (CGM) is broadly defined as the reservoir of baryonic material surrounding a galaxy~\citep[see][for a review]{tumlinson_2017}.  It holds material that serves as fuel for future star and black hole growth, as well as the by-products of feedback processes from star formation and active galactic nuclei (AGN) \citep[e.g.][]{hafen_2019}. A significant fraction of the metal budgets of $L\star$ galaxies lies in the CGM \citep{peeples_2014}. Additionally, the loss of gas from the CGM via AGN-driven `preventative' feedback has been linked to galaxy quenching and morphological transformation \citep{davies_crain_2020, oppenheimer_2020a}. The CGM is regarded as a key for understanding the cycle of baryons between galaxies and their surrounding environments \citep{hafen_2020} that is responsible for setting the physical properties of galaxies and their evolution over time \citep[see][for a review]{peroux_howk_2020}.

The CGM can be probed via its absorption or emission properties.  Owing to its diffuse and multi-phase nature, the CGM is typically very faint in emission \citep{putman_2012} except in the most massive halos where hot X-ray emitting gas is present \citep{werner_2020}. For more typical halos, absorption lines present the best approach to measuring CGM gas properties as they are sensitive to lower column densities. Absorption line studies require finding a coincident bright source (usually a quasar) illuminating a foreground CGM \citep[e.g.][]{tumlinson_2013, bordoloi_2014, borthakur_2015, berg_2018}, or alternatively quantifying the galaxy population along a given line of sight (LOS), and then measuring the desired atomic transitions at the redshift of the intervening CGM \citep[e.g.][]{rudie_2012}.  While this technique can probe fairly diffuse gas in multiple elements and ionic states, it is limited by having typically a single or small number of LOS probing a given CGM, and only in select transitions that are strong and in the observable wavelength window.  Since many of the strongest CGM transitions are in the ultraviolet (UV), at redshifts $z\la 1$ this typically requires space-based UV observations, adding to the challenge.

Hydrodynamic simulations provide a holistic approach to model the CGM of galaxies and help interpret absorption line observations.  By extracting spectra through the simulation volume, it is possible to compile statistics of simulated absorption lines and investigate the origin and underlying physical conditions of CGM absorbers. 
\cite{hani_2019} employed this approach using the Auriga simulations \citep{grand_2017} to highlight the wide diversity of physical conditions in the CGM of \lstar galaxies.
Such studies have revealed that galactic outflows are responsible for supplying the CGM with the enriched gas that is observed via metal absorption lines \citep{oppenheimer_2012, ford_2013, ford_2014, hummels_2013}, which is reflected in an angular dependence of CGM metallicity \citep{peroux_2020}.
Absorption from low metal ions arises from recently enriched, dense regions close to galaxies that will be re-accreted onto galaxies within a few Gyr; in contrast, high metal ions trace more diffuse gas that was enriched via outflows many Gyr ago \citep{oppenheimer_2009a, ford_2013, ford_2014, oppenheimer_2012}. 
The neutral hydrogen (\HI) absorption of moderate column density arises from gas that is metal poor, whereas strong \HI\ absorbers trace cool, enriched gas that is typically inflowing \citep{ford_2014}.
\cite{oppenheimer_2018b} compiled this picture into a two-phase model of the CGM around \lstar galaxies in which halo gas within $0.5 r_{200}$ is composed of cool, low metal line clouds within a hot ambient medium traced by higher metal lines, while halo gas outwith $0.5 r_{200}$ is heated to the virial temperature. 
Such cool gas clouds are thought to cool from initial local overdensities which trigger thermal instabilities \citep{nelson_2020}.

Careful comparisons between full cosmological galaxy evolution simulations and observations can provide constraints on models and highlight discrepancies between model predictions and the real universe. 
Early cosmological simulations used statistics of Ly-$\alpha$ absorption in the intergalactic medium (IGM) to constrain cosmologies \citep[e.g.][]{dave_1999}. 
More recent numerical studies have used contemporary IGM/CGM absorption observations as benchmarks to test the success of their simulations \citep[e.g.][]{dave_2010,  oppenheimer_2012, peroux_2020} and investigate the impact of physical processes such as turbulence \citep{oppenheimer_2009a} and stellar feedback \cite{hummels_2013}.
\cite{gutcke_2017} presented a comparison of the NIHAO simulation suite \citep{wang_2015} to \HI\ and \OVI\ absorption statistics, finding good agreement with observations.
\cite{ford_2016} tested models for stellar winds by comparing their simulations to the COS-Halos survey of \lstar galaxies \citep{tumlinson_2013, werk_2014}, finding that strong winds are necessary to reproduce CGM metal absorption.
The EAGLE simulations \citep{crain_2015, schaye_2015} have also been compared with COS-Halos \citep{oppenheimer_2016, oppenheimer_2018b}, showing that EAGLE reproduces the correlation between \OVI\ column density and star formation rate and is in good agreement with the low metal line statistics.
The IllustrisTNG simulations \citep{pillepich_2018} have likewise been shown to reproduce the observed dichotomy in \OVI\ absorption between star forming and quenched galaxies seen in COS-Halos \citep{nelson_2018b}, as well as reasonable \MgII\ absorber populations \citep{defelippis_2021}.

In \citet{appleby_2021} we explored the CGM as seen in absorption lines in the \simba\ simulation \citep{dave_2019}.  In particular, we tailored the extracted spectra to mimic the COS-Halos \citep{tumlinson_2013, werk_2014} and COS-Dwarfs \citep{bordoloi_2014} surveys, spanning three orders of magnitude in galaxy and halo mass.  We showed that the absorption line properties of \HI\ and selected metal ions (\MgII, \SiIII, \CIV, \OVI) around star forming galaxies are in good agreement with the observations, depending on the assumed photoionising background. In particular, \simba\ reproduces \HI\ absorption very well around star forming galaxies. We also examined the mass and metal budgets of halos in our simulations, and found that halo baryon fractions are generally less than half of the cosmic fraction, due to stellar feedback at low stellar masses and jet-mode AGN feedback at high stellar masses. Around quenched galaxies, the CGM baryons that remain are dominated by hot gas with $T >0.5T_{\rm vir}$, while the CGM of star-forming galaxies is more multi-phase.

Cosmological simulations enable the investigation of the CGM around a wide range of galaxy types.  However, they lack the resolution that modern zoom re-simulations can achieve, specifically in the diffuse CGM that is coarsely sampled by Lagrangian fluid elements.  Recently, several groups have presented ``CGM zoom'' simulations, where the resolving power is focused within the CGM rather than the galaxy. 
Such simulations have found that resolution is an important consideration, although their galaxy sample sizes are low due to the computational cost. For example, \cite{hummels_2019} demonstrated that increasing resolution in the halo leads to diminished warm/hot gas content and enhanced cool gas content, where the cool structures are able to survive for longer timescales and at smaller masses. This results in an increase in absorption of low ions such as \HI, and a decrease in absorption of high ions such as \OVI. 
Similarly, \citet{peeples_2019} and \citet{van_de_voort_2019} both find in their CGM zoom simulations that although the average CGM properties are unaffected, the distribution of physical conditions in the CGM are sensitive to resolution, which has consequences for \HI and metal absorption. \citet{van_de_voort_2019} find that this leads to enhanced \HI column densities in the outer CGM. Moreover, the typical CGM structures in simulations with enhanced resolution have lower masses down to $<10^4\msolar$, which is far below \simba's resolution limit. 
In the IllustrisTNG cosmological simulations, increased resolution has also been shown to lead to an increase in the formation of cold, small-scale CGM clouds \citep{nelson_2020}.
Numerical resolution clearly makes a difference for e.g. very high column density \HI\ absorbers.  However, \cite{suresh_2019} found using IllustrisTNG that metal absorption in the CGM was relatively insensitive to resolution.
Although the precise, quantitative impact of resolution on the current work is unclear, it is nonetheless an important consideration.

\simba's resolution is even lower than IllustrisTNG, which is a limitation of the current work. On the flip side, \simba provides large galaxy sample with a unique galaxy formation model that reproduces a wide variety of galaxy observations at low redshifts \citep{dave_2019,li_2019,thomas_2019,dave_2020,appleby_2020,thomas_2020, glowacki_2020, cui_2021} and low-redshift \lya\ statistics along random lines of sight \citep{christiansen_2020}, which make it valuable to examine despite its modest resolution. In higher resolution tests, \cite{appleby_2021} showed that increased resolution in \simba\ leads to enhanced absorption around star forming galaxies, for all ions probed across a range of ionisation energies. Hence it is likely that the absorber sample presented in this work represent minimum column densities and absorber counts. However, the same numerical tests also showed that simulation volume has an effect: the smaller boxes contain fewer of the most massive galaxies, leading to weaker preventative AGN feedback and enhanced CGM absorption. As such, the cosmological volumes offered by simulations such as \simba provide an important large-scale context for studying the CGM.

In this paper, we follow on our previous work to investigate the underlying physical nature of the population of CGM absorbers within \simba. We construct a new carefully-selected sample of synthetic spectra for a range of key ions as a function of both galaxy stellar mass and specific star formation rate (sSFR). In particular, we present counting statistics of the CGM absorber population and study properties of the absorbing CGM gas by examining its distribution within cosmic physical phase space. We study the relationship between physical density and column density of CGM absorbers, and investigate the contribution of collisional ionisation to the absorber statistics. Lastly, we quantify the contribution to the LOS spectra from satellite galaxies living in the halos of other galaxies. 

This paper is organised as follows. In \S\ref{sec:sims} we present the
\simba simulations. In \S\ref{sec:sample} we describe the galaxy selection, spectrum generation and fitting processes. In \S\ref{sec:absorbers} we present the CGM counting statistics and physical underlying conditions. In \S\ref{sec:redshift} we discuss the redshift dependence of our results. Finally in \S\ref{sec:satellites} we present the contributions of satellite galaxies to the original absorber sample.

\section{Simulations}\label{sec:sims}

% General
\simba\ \citep{dave_2019} is a suite of state-of-the-art cosmological simulations and is the successor to the earlier \mufasa\ simulations \citep{dave_2016}. \simba\ is run using a branched version of the gravity plus hydrodynamics code \gizmo\ \citep{hopkins_2015} in its Meshless Finite Mass (MFM) mode. \simba's fiducial simulation is a $100\ \hmpc$ volume evolved from $z=249$ to $z=0$, with $1024^3$ dark matter particles and gas elements. The mass resolution is $9.6\times 10^7\msolar$ and $1.82\times 10^7\msolar$, respectively. The minimum adaptive gravitational softening length is $\epsilon_{\rm min} = 0.5\hkpc$, corresponding to 0.5 times the mean spacing between dark matter particles. Cosmological initial conditions are generated using \music\ \citep{hahn_2011} assuming a \cite{planck_collab_2016} cosmology with $\Omega_m = 0.3$, $\Omega_b = 0.048$, $\Omega_\Lambda = 0.7$, $H_0 = 68\ {\rm km\ s}^{-1}\mpc$, $\sigma_8 = 0.82$ and $n_s = 0.97$.

% Cooling and heating
Photoionisation heating and radiative cooling are implemented using the \grackle-3.1 library \citep{smith_2017} in its non-equilibrium mode. The neutral hydrogen fraction of gas particles is computed self-consistently throughout the simulation, accounting for self-shielding following the \cite{rahmati_2013a} prescription, whereby the metagalactic ionising flux is attenuated based on gas density. A spatially uniform \cite{haardt_2012} ionising background is assumed. 

% Star formation
Star formation is modelled using an H$_2$-based \cite{schmidt_1959} law, with ${\rm SFR} = 0.02\rho_{H_2}/t_{\rm dyn}$. The H$_2$ content of the gas particles is computed based on their metallicity and local column density, using the sub-grid \cite{krumholz_2011} prescription. Artificial interstellar medium (ISM) pressurisation is included at the minimum level required to resolve the Jeans mass in star-forming gas \citep{dave_2016}.

% Chemical enrichment
The chemical enrichment model tracks 11 elements (H, He,  C, N, O, Ne, Mg, Si, S, Ca, Fe) during the simulation, from Type II supernovae (SNII), Type Ia supernovae (SN1a) and asymptotic giant branch (AGB) stars, as described in \cite{oppenheimer_2006}. The yield tables employed are that of \cite{nomoto_2006} for SNII yields, \cite{iwamoto_1999} for SN1a (assuming a metal yield of 1.4$\msolar$ per supernova), and \cite{oppenheimer_2006} for AGB yields (assuming a helium fraction of 36\%, a nitrogen yield of 0.00118, and a \cite{chabrier_2003} initial mass function). 
Cosmic dust is tracked as a fraction each gas element's metal mass that is passively advected along with the gas, described in \cite{li_2019} and broadly following \cite{mckinnon_2016}. The dust particles are grown via condensation and accretion of gas-phase metals via two-body collisions; conversely they can be destroyed via supernova shocks and collisions with thermally excited gas elements. Mass and metals from destroyed dust particles are returned to the gaseous phase.

% Star formation driven winds
Star formation driven-winds are included as two-phase, decoupled, metal-enriched winds, with 30\% of the wind particles ejected hot and the rest ejected at $T\approx 10^3$K.  The mass loading factor scales with $\mstar$ based on mass outflow rates from the Feedback In Realistic Environments \citep[FIRE,][]{hopkins_2014} simulations \citep{angles-alcazar_2017b}, where $\mstar$ is computed with an on-the-fly friends-of-friends finder applied to stars and ISM gas. The wind velocity scaling is also based on results from the FIRE simulations \citep{muratov_2015}, modified to account for hydrodynamic slowing and the difference in launch site between \simba\ and FIRE. Ejected wind particles are temporarily hydrodynamically decoupled to avoid numerical inaccuracies, with cooling paused to allow the winds to deposit thermal energy into the CGM. Wind particles are enriched with metals from their surroundings at launch time, modelling metal loading by SNII. The star formation-driven wind mechanisms are described in more detail in \cite{appleby_2021}.

% Black hole growth
\simba\ employs a method for black hole growth and feedback that is novel amongst cosmological simulations. Black holes are seeded with $M_{\rm seed} = 10^4 \msolar$ into galaxies with $\mstar \gtrsim 10^{9.5} \msolar$ and grown self-consistently during the simulation via a two-mode accretion prescription. For cold gas (T $< 10^5$\ K) surrounding the black hole, torque-limited accretion is implemented following the prescription of  \cite{angles-alcazar_2017a}, which is based on \cite{hopkins_2011a}. For hot gas (T $> 10^5$\ K) classical Bondi accretion is adopted.

% Black hole feedback
Kinetic AGN feedback is implemented in two modes: `radiative' and `jet' at high and low Eddington ratios ($\fedd$), respectively. This is designed to mimic the dichotomy in black hole growth modes observed in real AGN \citep[e.g.][]{heckman_2014}. Radiative and jet feedback outflows are ejected in a direction $\pm$ the angular momentum of the inner disk and with zero opening angle. The mass outflow rate varies such that the momentum output of the black hole is $20L/c$, where $L$ is the bolometric AGN luminosity. In the radiative mode, winds are ejected at the ISM temperature with a black hole mass-dependent velocity, calibrated using X-ray detected AGN in SDSS \citep{perna_2017}. At $\fedd < 0.2$ the transition to jet mode begins for galaxies with $M_{\rm BH} > 10^{7.5}M_{\odot}$, with winds ejected at the halo virial temperature. Jet outflows receive a velocity boost that increases with decreasing $\fedd$, up to a maximum of 7000 km s$^{-1}$ above the radiative mode velocity at $\fedd = 0.02$. A third mode of AGN feedback is implemented to mimic X-ray heating from the black hole accretion disk, following \cite{choi_2012}. X-ray feedback only occurs for galaxies with $f_{\rm gas} < 0.2$ and with full-velocity jets. A suite of feedback variant \simba\ simulations exist to test the impact of the feedback modes; however in this work we adopt the fiducial \simba\ run with full feedback physics.

% Galaxy finder
Halos are identified during the simulation using {\sc Subfind}, which is implemented in \gizmo. Galaxies are identified in post-processing using a 6D friends-of-friends galaxy finder using all stars, black holes, and gas elements with $n_{\rm H}>n_{\rm th}$. We assume a spatial linking length of 0.0056 times the mean interparticle spacing (twice the minimum softening), and the velocity linking length is set by the local dispersion. Halos and galaxies are cross-matched using the {\tt yt}-based \citep{turk_2011} python package \caesar\footnote{\url{https://caesar.readthedocs.io}}. \caesar also pre-computes a number of global galaxy properties, such as stellar mass and the instantaneous star formation rate (SFR). Particle data is read using \textsc{PyGadgetReader}\footnote{\url{https://github.com/jveitchmichaelis/pygadgetreader}}. 

\section{Absorber sample}\label{sec:sample}

\subsection{Galaxy selection}

A representative sample of galaxies is obtained across a range of physical properties within each of the $z =$ 0, 0.25, 0.5 and 1 snapshots, by defining regions in $\mstar$-sSFR space and selecting galaxies within each region. We define three categories of galaxies within each mass bin: 1) star-forming galaxies, with ${\rm log (sSFR/Gyr}^{-1}) > -1.8 +0.3z$ for consistency with previous work with the \simba\ simulation \citep[e.g.][]{thomas_2019}; 2) green valley galaxies, with sSFR within 1 dex below the star-forming galaxy threshold; 3) quenched galaxies, with SFR $= 0\ {\rm M}_{\odot} {\rm yr}^{-1}$ to ensure that these galaxies are fully quenched. We impose a lower stellar mass limit of $\mstar > 10^{10}\msolar$ and require that the galaxies are the central galaxy in their halos. We select 12 galaxies in each $\mstar$-SFR region (apart from in the highest mass green valley region, in which there are only 8 galaxies matching these criteria at $z=0$). Subsamples of 12 were chosen to balance a reasonable sample size with the computational cost of generating spectra and the size of the smallest $\mstar$-SFR bin. All other bins contain more than 12 galaxies; in these bins we randomly select galaxies using {\sc Numpy}'s {\tt random.choice} function with a seed of 1 and assuming a uniform distribution. We select a constant number of galaxies per bin to ensure that any surplus or deficit of CGM absorbers around specific types of galaxies is not due to the over-representation of such galaxies in our initial sample. We have verified that the galaxies in our sample are randomly orientated with respect to the line of sight and thus our results are not biased due to galaxy inclination.

\begin{figure}
	\includegraphics[width=\columnwidth]{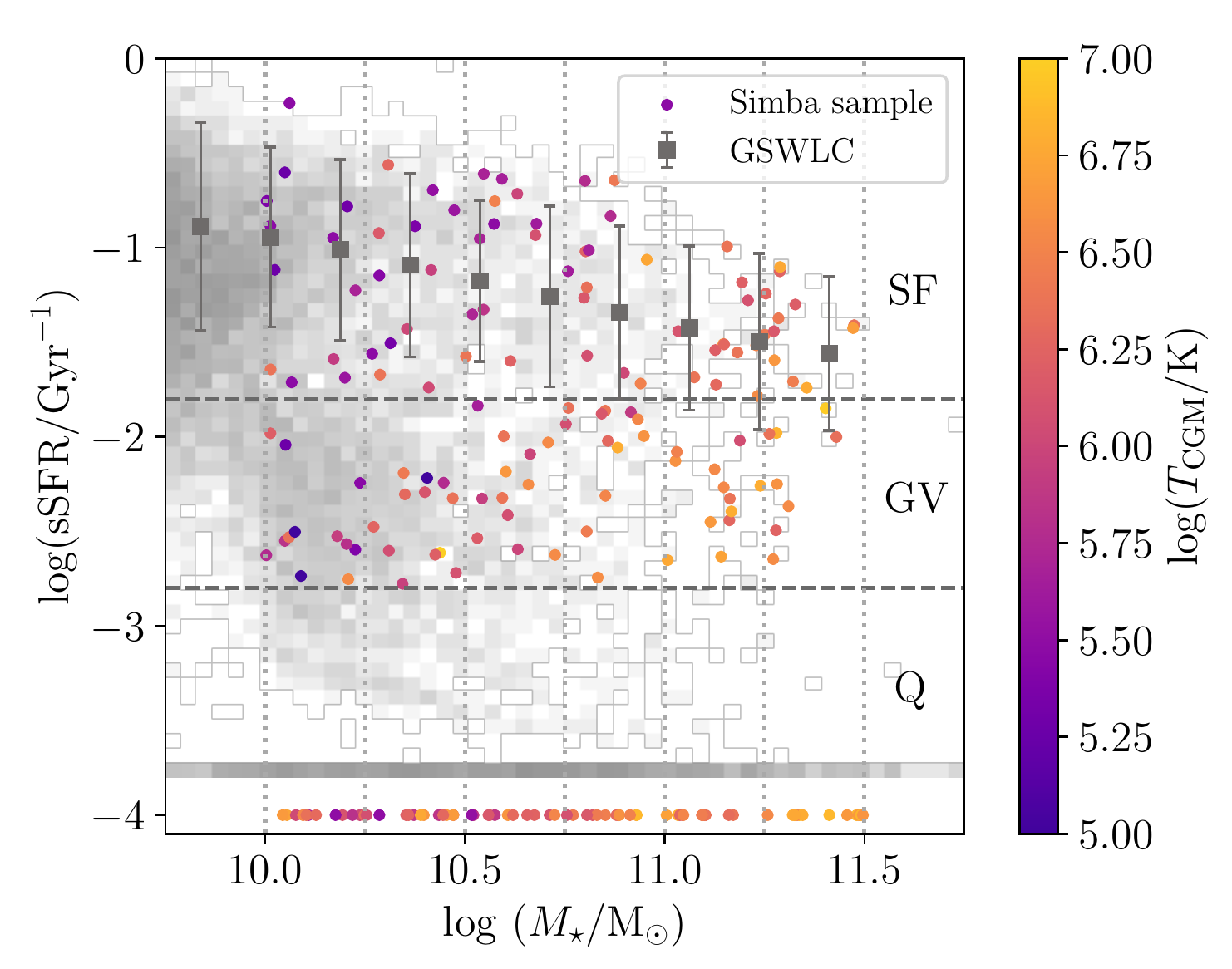}
	\vskip-0.1in
    \caption{Specific star formation rate (sSFR) against stellar mass ($\mstar$) for our $z=0$ galaxy sample, colour-coded by the mass-weighted average halo temperature. Horizontal lines of constant sSFR indicate the partition between star-forming, green valley and quenched galaxies. Vertical lines show the boundaries between \mstar\ bins. The observed star-forming main sequence from GSWLC is shown as black squares. The \simba galaxy population as a whole are shown as a log-scale 2D histogram in greyscale.} 
    \label{fig:galaxy_sample}
\end{figure}

Figure \ref{fig:galaxy_sample} shows sSFR against $\mstar$ for our $z = 0$ galaxy sample. Horizontal lines of constant sSFR show our boundaries between star-forming, green valley and quenched galaxies. The vertical lines indicate the boundaries between our stellar mass bins. The color scale shows the mass-weighted average temperature of gas elements in each galaxy's host halo. We show how this sample relates to the overall \simba population with a 2D log-scale histogram of \simba\ galaxies in greyscale. Galaxies with zero star formation are plotted at ${\rm sSFR} = 10^{-3.8}{\rm Gyr}$. A comparison with the observed star-forming main sequence of galaxies (${\rm sSFR} > 10^{-1.8}{\rm Gyr}$) from the GALEX-SDSS-WISE Legacy Catalog \citep[GSWLC,][]{salim_2016, salim_2018} is shown as a running median in black squares. The error bars are the standard deviation within each bin. The overall population of star forming \simba galaxies at low redshift reproduces the observed main sequence well \citep{dave_2019}.  

\subsection{Spectral generation}\label{sec:spectra}

For each galaxy in our sample, we probe its CGM by generating mock spectra for multiple lines of sight (LOS) through the simulation box. We select impact parameters ($r_\perp$) for each galaxy in multiples of the halo $r_{200}$, in order to probe the impact of radial distance from the galaxy whilst placing galaxies of varying masses on a common scale. For each impact parameter, we select 8 equally-spaced LOS in a circle around the galaxy with radius $r_\perp$, starting from ($x_{\rm gal}+r_\perp, y_{\rm gal}$) and proceeding anti-clockwise. This results in 8480 LOS spectra per line transition. Absorption spectra are generated for a selection of commonly observed ions spanning a range of excitation energies: \HI\ 1215\AA, \MgII\ 2796\AA, \CII\ 1334\AA, \SiIII\ 1206\AA, \CIV\ 1548\AA\ and \OVI\ 1031\AA.

Mock spectra are generated following the same procedure as in \cite{appleby_2021}. The python analysis package \pygad\ \citep{rottgers_2020} is used to generate absorption spectra through the simulation volume in the $z$-axis direction by identifying the gas elements whose smoothing lengths intersect with the given LOS. For each gas element, the ionisation fractions are found using look up tables generated with version 17.01 of the \cloudy\ cloud simulation code \citep{ferland_2017} using Cloudy Cooling Tools\footnote{\url{https://github.com/brittonsmith/cloudy\_cooling\_tools}}. Computing ionisation fractions with \cloudy\ requires an assumption for the incident photoionising background spectrum; indeed earlier studies of the CGM in \simba\ demonstrate that the choice of photoionising background has a significant impact on the metal line absorption statistics \citep{appleby_2021}. For consistency with earlier work on the CGM and IGM in \simba\ \citep{christiansen_2020, appleby_2021, bradley_2022}, we assume a \cite{faucher-giguere_2020} photoionising UV background. For neutral \HI, self-shielding is already applied directly during the simulation run; for metal lines, self-shielding is applied following the \cite{rahmati_2013a} prescription, whereby an overall attenuation factor is applied to the entire ionising background spectrum in the optically thin limit. 

The ion densities for each gas element are found by multiplying the gas densities by the ionisation fractions for each species, assuming a $\sim$76\% mass fraction for hydrogen and the individually-tracked mass fractions within \simba\ for the metals. The ion densities are smoothed along the LOS into pixels using the gas elements' individual smoothing lengths and metal masses (for metal lines), and using the same spline kernel used in the \gizmo simulation code. The spectral pixel scale is $2.5{\rm km\ s}^{-1}$. The optical depths for each pixel are computed from the resulting ion column densities using the oscillator strength for each species. Column density-weighted physical density, temperature, metallicity, and peculiar velocity are also computed in the same manner within the LOS pixels. For further details on \pygad,\ see \citet{rottgers_2020}. We exclude wind particles from the spectral generation since these gas elements are hydrodynamically decoupled from the surrounding gas; these represent a very small fraction of CGM mass in the halos of the galaxies we consider~\citep{appleby_2021}.

We apply the following effects to the resulting synthetic spectra to mimic observations gathered with Hubble Space Telescope's Space Telescope Imaging Spectrograph:
\begin{enumerate}
    \item The spectra are convolved with the STIS E140M line spread function (LSF), which is the medium resolution Multi-Anode Microchannel Array (MAMA) Echelle spectrograph LSF, centered on 140nm.  The typical FWHM is around 6${\rm km\ s}^{-1}$, while the pixel scale is typically 2.5${\rm km\ s}^{-1}$. For \MgII, we convolve with a Gaussian with FWHM of 6${\rm km\ s}^{-1}$ to mimic Keck spectra.
    \item Gaussian random noise is added to the spectra with a signal to noise ratio (SNR) of 30.
    \item A mock continuum fitting procedure is applied to the spectra as in \cite{appleby_2021} (iteratively removing pixels that are $> 2\sigma$ below a polynomial best-fit to the spectrum until the fractional change between continuum fits is less than $10^{-4}$).
\end{enumerate} 
While our goal is not to compare with observations, this choice allows us to examine absorber properties at the highest resolution among currently available UV spectrographs capable of obtaining significant samples of CGM absorbers.  The resolution of the COS spectrograph, in contrast, is often insufficient to resolve metal lines arising in cooler gas, which would make Voigt profile fitting parameters (described in the next section) less robust.

\begin{figure*}
    \includegraphics[width=\textwidth]{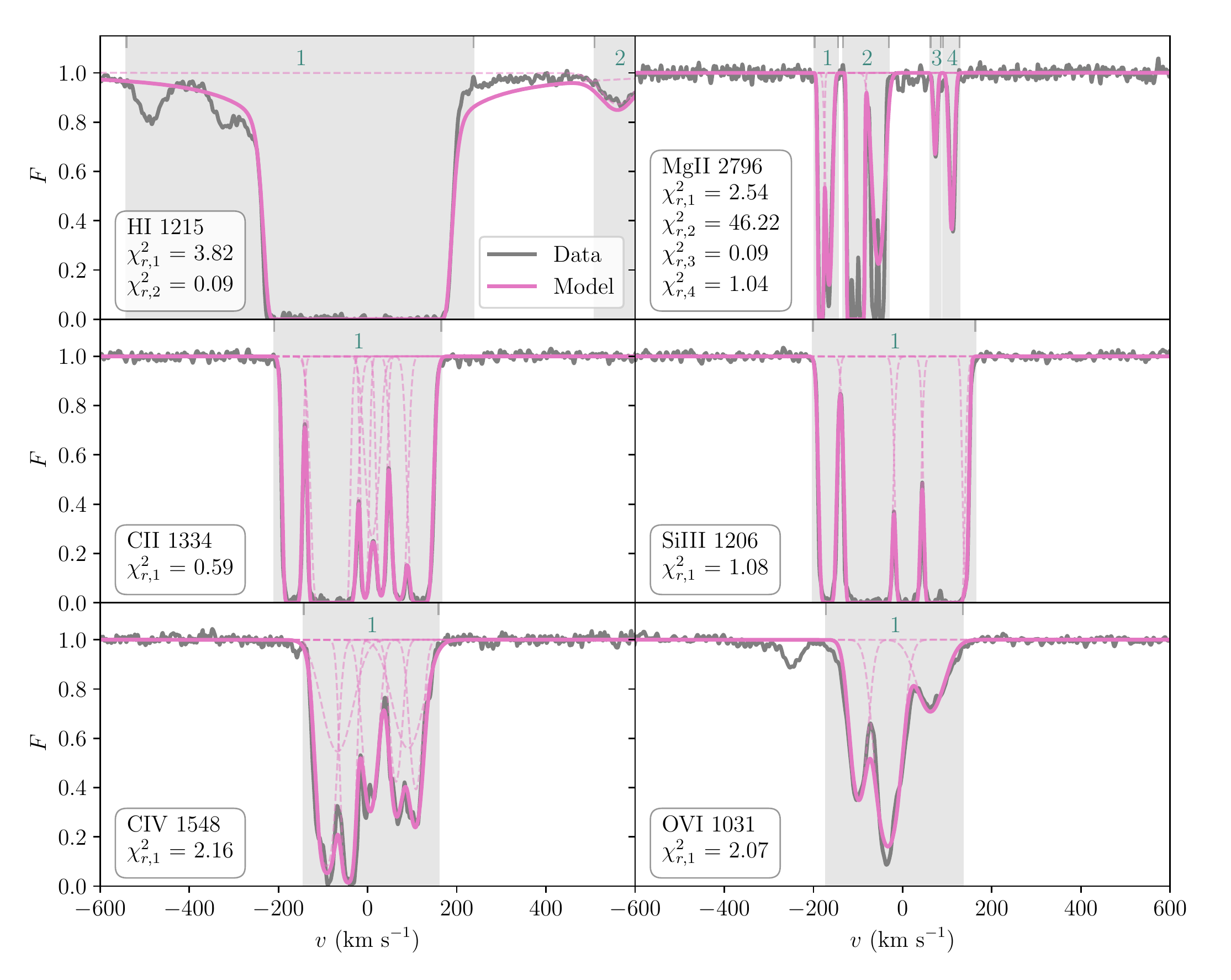}
	\vskip-0.1in
    \caption{Example synthetic spectra (grey lines) of the CGM of a star forming galaxy from our $z=0$ sample with $\mstar = 10^{10.8}\msolar$ and ${\rm SFR} = 14.2\msolar {\rm yr}^{-1}$. The line of sight is $r_\perp = 0.25r_{200} = 155\ \hkpc$ from the galaxy. The spectra are centered on the galaxy's systematic velocity. The detected absorption regions are shaded in grey and numbered at the top of each plot. The solid pink lines show the overall fitted model; where models consist of multiple lines, these are shown as dashed pink lines. The $\chi_r^2$ for each region is given in the legend on the lower left of each panel.}
    \label{fig:example_spectrum}
\end{figure*}

Figure \ref{fig:example_spectrum} shows example synthetic CGM absorption spectra from our sample (grey lines). The central galaxy is a star forming $\mstar = 10^{10.8}\msolar$, ${\rm SFR} = 14.2\msolar {\rm yr}^{-1}$ galaxy at $z=0$. Lower mass, star forming galaxies such as this are over-represented in our absorber sample since their CGM contain an over-abundance of absorbers compared with the population as a whole (\S\ref{sec:cddf}). The impact parameter of the line of sight is $r_\perp = 0.25r_{200} = 155\ \hkpc$ and the spectrum is centered on the galaxy systematic velocity. Each panel shows the spectrum for a different ion.
The shapes of these spectra are representative of our spectrum sample:
the \HI\ spectrum shows strong, saturated absorption with broad wings, while the metal line spectra also show strong absorption, with \MgII, \CII\ and \SiIII\ also saturating in places.
The \MgII\ spectrum is composed of many narrow lines, which we find is typical of \MgII\ absorption in \simba.
The \HI\ and metal absorption lines appear aligned with one another in velocity space, despite the range of physical conditions probed with these lines.

\subsection{Spectrum fitting}\label{sec:fitting}

We extract observables from the spectra within a $\pm 600 \kms$ window centered on each galaxy. Fitting Voigt profiles to the spectra provides the column density $N$, the Doppler $b$ parameter, the wavelength (or velocity) location along the LOS, and equivalent width (EW) for each line in the absorption system. We also compute the EW directly from the flux for comparison.

The Voigt profile fitting procedure broadly follows that of {\sc AutoVP} \citep{dave_1997}, but reformulated into Python and integrated into {\sc Pygad}. To simplify the fitting procedure, we first identify absorption regions by computing the detection significance ratio of each pixel, which is defined as the Gaussian-smoothed flux equivalent width divided by the Gaussian-smoothed noise equivalent width. Absorption regions are identified where the flux drops below the continuum level and the ratio is $>5\sigma$. We ensure that the region edges begin at continuum level, and merge together nearby regions within 2 pixels of one another. We impose a region length limit of $>2$ pixels to avoid spurious absorbers. Table \ref{tab:absorber_sample} gives the number of LOS (out of the 8480 LOS per species) that have no detected absorption regions.

Each absorption region is fitted individually. In the first instance a line is added at the position of lowest flux, and an initial estimate for $N$ and $b$ is made based on the depth and velocity width of the local flux minimum, respectively. For saturated lines (those with a residual flux below the noise level), a line is placed in the middle of the saturated trough and the initial estimate for $N$ and $b$ is made by finding the values that give the lowest reduced $\chi_r^2$ in a coarse grid of log$N$-log$b$ space. The best-fit Voigt profile solution is then found using the {\tt minimize} routine from the {\tt scipy.optimize} subpackage\footnote{\url{https://docs.scipy.org/doc/scipy/reference/optimize.html}} to search parameter space for the line parameters which minimise $\chi_r^2$. We supply prior bounds of $12 <{\rm log}(N/{\rm cm}^{-2})<19$ for \HI\ and $11 <{\rm log}(N/{\rm cm}^{-2})<17$ for metal lines. Bounds on $b$ are computed based on the thermal line width at $10^4$K and $10^7$K. To avoid overfitting that occurs with steeply saturated absorbers, the $\chi_r^2$ is asymmetrically penalised further where the model is $>3\sigma$ below the data. 

If $\chi_r^2 < 2.5$ is reached then the solution is accepted, as we find that this gives a reasonable visual fit to the data in most cases. Otherwise, the model is subtracted from the data to find a residual flux and a line is placed at the residual minimum, and the process is repeated until an acceptable fit is achieved, up to a maximum of 10 lines. New lines must improve the $\chi_r^2$ by at least 5\%, and if the model does not improve sufficiently after 2 additional lines then the process is halted, and the additional lines are ignored. If after 10 lines an acceptable solution is not found then we revert to the number of lines that performed best. Our routine then attempts to find the solution with the fewest lines by iteratively re-fitting with each line removed one at a time. If the $\chi_r^2$ acceptance threshold is reached, or the $\chi_r^2$ increases by less than 5\% then the line is removed from the solution. Formal errors in $N$ and $b$ are found by computing the Hessian covariance matrix. In cases where an acceptable fit is not found, the routine flags these for possible manual fitting, although for this work we do not revisit these fits. Table \ref{tab:absorber_sample} gives the number and fraction of fitted LOS that are flagged as having at least one absorption region with an unacceptable fit ($\chi_r^2> 50)$. We verify that the EWs resulting from these fits closely correspond to the EWs computed directly from the spectra, broadly following a 1:1 relation.

Although we adopt a $\chi_r^2 <2.5$ acceptance limit during the fitting process, the solution for a LOS does not always reach this threshold. To cope with such cases, we adopt $\chi_r^2$ acceptance thresholds for our absorber sample such that we recover 90\% of the total EW across all LOS for each species. While the Voigt profile fitter occasionally overestimates the EW for individual lines of sight, we have checked that the total EW computed directly by summing all lines of sight is always higher than the total EW obtained by fitting the spectra. The upper limits ($\chi_{r,90}^2$) are given in Table \ref{tab:absorber_sample}, along with the sample size and $N$ completeness limit for each species. In practice the quality of the fit is much better than these limits in most cases; the median $\chi_r^2$ of absorption lines in our sample is also given in Table \ref{tab:absorber_sample}. We are reluctant to adopt a stringent fixed $\chi_r^2$ for all species as this leads to an incomplete sample of absorbers. We show a comparison of individual LOS EW and $N$ obtained from our Voigt profile fitting routine to those obtained directly from the spectra in Appendix \ref{sec:appendix}.

Figure \ref{fig:example_spectrum} shows the Voigt profile fits for the example spectrum. The detected regions of absorption are shaded in grey and numbered along the top of each panel. The individual fitted absorber components are plotted as dashed pink lines, with the overall best-fit model plotted in solid pink. The $\chi_r^2$ for each absorption region is displayed in the lower left of each panel. Overall, the quality of the Voigt profile fits to the data are good considering the fits are produced automatically with no manual intervention. Inspecting the fits by eye, the models clearly capture the overall shape of the original spectra. 

The \MgII\ spectrum contains an example of a poor Voigt profile fit (absorption region 2 in the \MgII\ panel, with $\chi_r^2 = 46.22$). In this case the fitting routine has not succeeded in finding an accurate solution. However, by eye the fit is not completely unreasonable: the fit has two strong absorption features at the position of the two main absorption features in the data, but the fit has not captured the fine detail of the many small absorption lines within these features. This is typical of the high $\chi_r^2$ \MgII and \SiIII lines in our sample. 

\begin{table*}
    \centering
    \begin{tabular}{c|c|c|c|c|c|c|c|c}
        \hline
        Species & $E$(eV) & $n_{\rm LOS\ no-abs}$ & $n_{\rm reject}$ & $f_{\rm reject}$ & $n_{\rm absorbers}$ & ${\rm log }(N_{\rm min}/ {\rm cm}^{-2})$ &$\chi_r^{2,90}$ & Median $\chi_r^2$ \\
        \hline
        \HI & 13.60 & 2784 & 447 & 0.078 & 9664 & 12.7 & 4.0 & 1.2 \\
        \MgII & 15.04 & 7396 & 437 & 0.403 & 2911 & 11.5 & 50.0 & 2.2 \\
        \CII & 24.38 & 6195 & 393 & 0.172 & 5757 & 12.8 & 15.8 & 1.7 \\
        \SiIII & 33.49 & 5662 & 850 & 0.302 & 8097 & 11.7 & 39.8 & 3.0 \\
        \CIV & 64.49 & 4225 & 680 & 0.160 & 10752 & 12.8 & 8.9 & 1.7 \\
        \OVI & 138.12 & 3752 & 248 & 0.052 & 7986 & 13.2 & 4.5 & 1.6 \\
        \hline
    \end{tabular}
    \caption{Absorber sample properties: $E$, the excitation energy of the species; $n_{\rm LOS\ no-abs}$, the number of LOS with no detected absorption regions; $n_{\rm reject}$ and $f_{\rm reject}$, the number and fraction of fitted LOS flagged as having at least one unacceptable absorption region; $n_{\rm absorbers}$, the resulting number of fitted absorbers below the $\chi_r^2$ limit; ${\rm log }(N_{\rm min}/ {\rm cm}^{-2})$, the column density completeness limit; $\chi_r^{2,90}$, the $\chi_r^2$ below which we recover 90\% of the total EW; Median $\chi_r^2$, the median $\chi_r^2$ of all absorbers.}
    \label{tab:absorber_sample}
\end{table*}

\section{CGM absorbers}\label{sec:absorbers}

\subsection{Column density distribution functions}\label{sec:cddf}

We begin our analysis with counting statistics of the CGM absorber sample using the column density distribution function (CDDF), which gives the column density probability distribution normalised by the comoving path length along the LOS. The CDDF is defined as:
\begin{equation}
    f(N, z) = \frac{\partial^2n}{\partial N\partial X},
\end{equation}
where $n$ is the number of absorbers in each bin of ${\rm log}N$, and $\partial X$ is the comoving path length, designed to account for cosmological evolution \citep[e.g.][]{wijers_2019}:
\begin{equation}
    \partial X = {\rm d}z\frac{H_0}{H(z)}(1+z)^2,
\end{equation}
where ${\rm d}z=600\ {\rm km\ s}^{-1}\times N_{\rm spec}/c$ is the redshift path length (with $N_{\rm spec}$ being the number of spectra generated) and $H(z)$ is the Hubble constant at $z$.

\begin{figure*}
	\includegraphics[width=\textwidth]{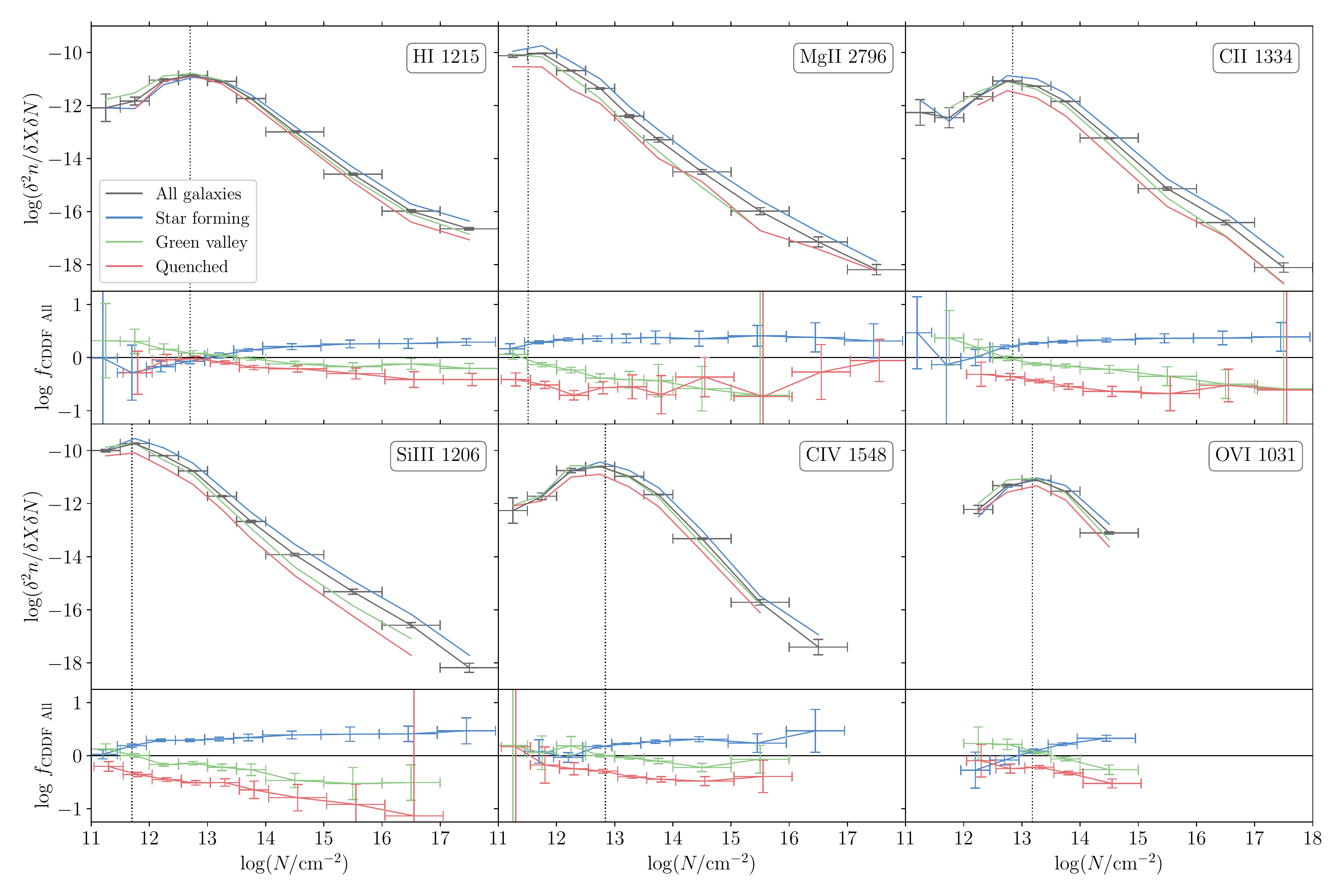}
	\vskip-0.1in
    \caption{The $z=0$ CDDF for each ion considered in this work. Colours represent galaxy types: all galaxies (grey), star forming (blue), green valley (green) and quenched (red). The smaller panels show the fraction with respect to the CDDF for all galaxies and at all impact parameters. Horizontal error bars show the width of the column density bins. Vertical error bars on show representative cosmic variance uncertainties, computed using all absorbers. Vertical lines indicate the completeness limit in each case.} 
    \label{fig:cddf}
\end{figure*}

Figure \ref{fig:cddf} shows \simba's predictions for the CDDF in the CGM for each species. Grey lines show the CDDF for all galaxies, while the blue, green, and red lines show the galaxies separated into star forming, green valley, and quenched sub-samples, respectively. The smaller panels show the CDDF fraction with respect to the case for all galaxies and at all impact parameters. Horizontal error bars show the width of the column density bins, where we have increased the bin width at the high column density end to account for low numbers. Vertical error bars show representative cosmic variance uncertainties, combined with the Poisson error within each bin, computed using all galaxies and all LOS. Error bars in the smaller panels are slightly offset for easier viewing. To determine the cosmic variance error, the $x$ and $y$ sides of the simulation box are each split into quarters, forming 16 equal-sized cells across the face of the simulation. The absorbers are binned into these cells based on the $x-y$ positions of their sightlines. The cosmic variance uncertainties are computed by jackknife resampling: we iterate over the cells and omit the absorbers from each cell in turn, and compute the CDDF from the remaining absorbers. The cosmic variance error is the variance of the 16 estimates of the CDDF. The uncertainties in the smaller panels are those of from all galaxies, added in quadrature with the uncertainties from star-forming, green valley and quenched galaxies. Where the uncertainties are large (e.g. at the high column density end for \CII and \SiIII), this arises from low numbers of such absorbers.

The overall shape of the CDDF in each case is that of a power law at high column densities with a turnover at the low column end around ${\rm log} (N / {\rm cm}^{-2}) > 12-13 $ (the exact turnover column density varies for each ion). The turnover may be due to incompleteness in our sample of low column density absorbers, or a genuine physical decrease in absorbers of this strength. We have conducted completeness tests on our sample by re-generating our LOS sample using an SNR of 100 to see whether this uncovers low column density absorbers that otherwise would be below the noise limit. We see little difference in the CDDFs from this high SNR sample, suggesting that such low column density absorbers genuinely do not exist in our simulation, likely due to numerical resolution. We compute lower completeness limits on our sample by fitting the power law portion of each CDDF and identifying where the CDDF falls below 50\% of the expectation at low column densities. The completeness limits for each species are shown as dashed vertical lines in Figure \ref{fig:cddf} and given in Table \ref{tab:absorber_sample}. For the remainder of this work we will only discuss results for absorbers above these limits.
%We conclude that our absorber sample is complete within the limitations of the simulation resolution. 

% HI
Considering the CDDFs for absorbers from all galaxies in our sample, we find abundant \HI\ absorption in our CGM sample, even out to high column densities, resulting in a shallow power law slope for \HI.
% MgII
\MgII absorbers are the least frequently occuring ion in our sample, and typically have low column density. Hence the CDDF for \MgII\ barely probes the low column density turnover.
% CII
% SiIII
% CIV
% OVI
Our absorber sample does not contain any \OVI\ absorbers outside of a narrow range of column densities ($12 < {\rm log} (N /{\rm cm}^{-2} < 15$), resulting in a CDDF which probes only the turnover and a limited portion of the power law. This reflects our spectral generation approach, which provides limited sampling at very high column densities. In contrast, \cite{bradley_2022} also used \simba\ to examine global \OVI\ statistics along random lines of sight using projected column density maps through the simulation volume, which allowed many more lines of sight thus enabling the \OVI\ CDDF to be measured to higher column densities; they showed that the resulting CDDF is in very good agreement with observations at $z\la 0.7$.

% SF vs GV vs Q
Splitting the absorbers by the star formation activity of the central galaxy, we can begin to see the effect of star formation on the CGM. For each ion, absorbers from around star forming galaxies have higher number counts at all column densities. Thus star forming galaxies contribute more to the overall CDDF. For \SiIII\ and \CIV, the high column density absorbers only occur in the CGM of star forming galaxies. The difference plots show the relative contribution of each galaxy subsample to the total CDDF; for each species, the difference plots show a positive (negative) slope for star forming (green valley and quenched) galaxies. 

Examining the absorber statistics from the green valley sample, is it clear that the population of CGM absorbers around green valley galaxies more closely resembles that of quenched galaxies, rather than the star forming galaxies. This is particularly true for low ions, which (as we will later show) trace cooler, denser gas.  Although these green valley galaxies have some residual star formation, their CGM gas has already dropped out of the necessary conditions for abundant and high column density absorption. This suggests that the CGM of green valley galaxies `quenches' (that is, resembles that of a quenched galaxy) before star formation is fully halted in the central galaxy, at least in terms of their cool, dense gas content.

\begin{figure*}
	\includegraphics[width=\textwidth]{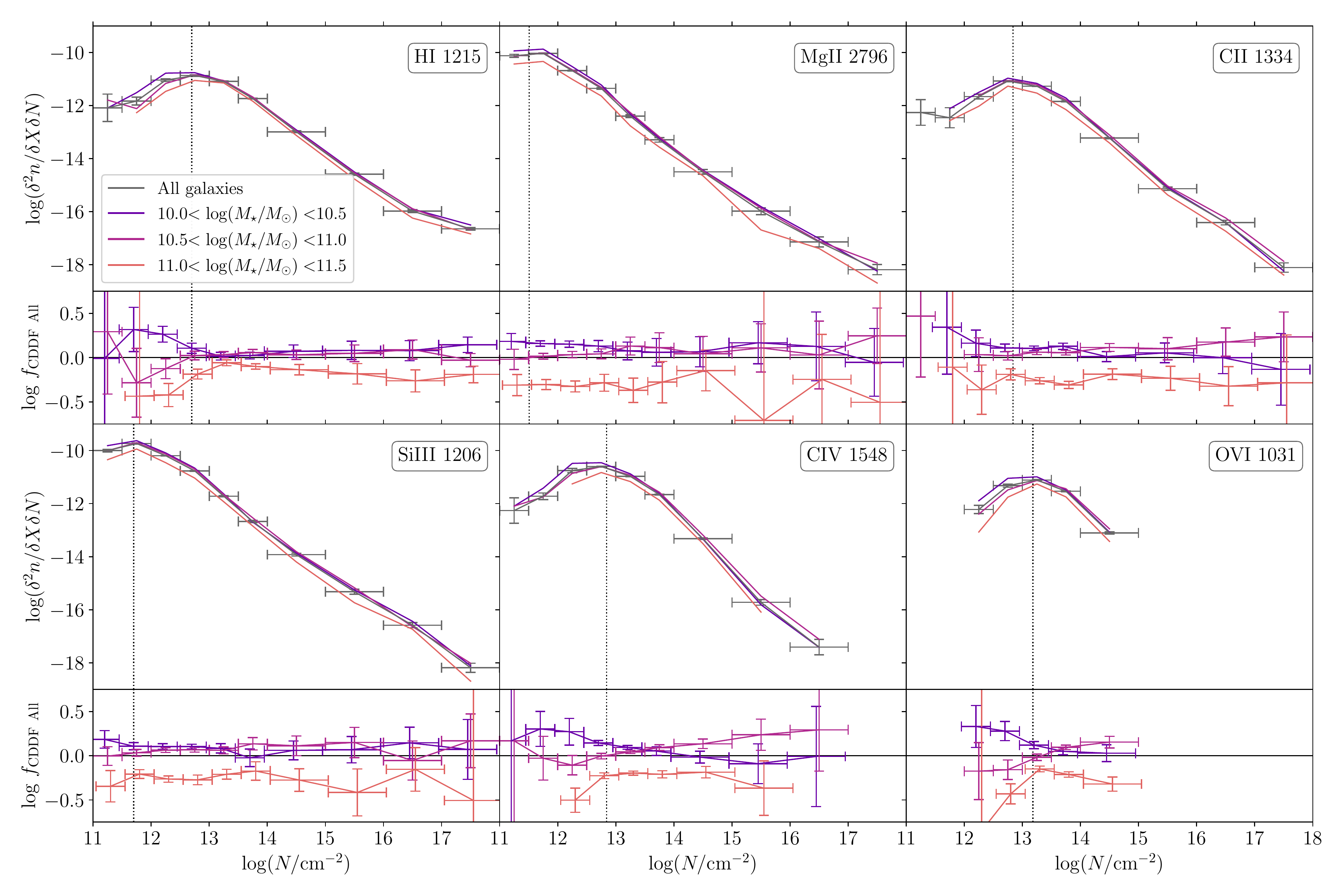}
	\vskip-0.1in
    \caption{As in Figure \ref{fig:cddf}, the $z=0$ CDDF for each ion considered in this work. In this case the colours represent broad stellar mass bins. The smaller panels show the fraction with respect to the CDDF for all galaxies and at all impact parameters. Horizontal error bars show the width of the column density bins. Vertical error bars on show representative cosmic variance uncertainties, computed using all absorbers. Vertical lines indicate the completeness limit in each case. } 
    \label{fig:cddf_mass}
\end{figure*}

Having examined the dependence on star formation activity, we now investigate the stellar mass dependence of \simba's absorber counting statistics. Figure \ref{fig:cddf_mass} again shows \simba's predictions for the CDDFs in the CGM; in this case the grey lines still represent the CDDF for all galaxies, while the blue, purple, and orange lines show the galaxies separated into broad stellar mass bins. As before, the smaller panels show the CDDF fraction with respect to the case for all galaxies at all impact parameters. The errorbars are computed in the same manner as Figure \ref{fig:cddf}. The vertical lines in each panel indicate the completeness limits. 

In general, there is little difference in the CDDF from galaxies in our low and intermediate mass bins, and these CDDFs essentially follow the shape and normalisation of the CDDF for the full galaxy sample. Where the CDDF from these stellar mass bins is substantially higher than the CDDF for all galaxies, this generally occurs below the completeness limit. Thus, the counting statistics of galaxies with $10 < {\rm log} (\mstar / \msolar) < 11$ are largely independence of stellar mass. In \cite{appleby_2021} we examined the mass and metal budgets of halos in \simba, and found that (when combining systems regardless of star formation activity) galaxies in this mass range have a multiphase CGM with roughly equal proportions of metal mass in cool, warm and hot CGM, as well as 10-20\% of metal mass gas in the ISM. Hence these galaxies have an abundance of metal enriched gas in their halos that is capable of producing measurable absorption.

However, galaxies in the highest mass bin ($11 < {\rm log} (\mstar / \msolar) < 11.5$) have significantly fewer absorbers at all column densities. The CDDF fraction from high mass galaxies is roughly constant with column density (above the completeness limit). The CDDF of \HI\ absorbers is at the level of $\sim70\%$ of the overall CDDF; for metal absorption it is slightly lower at $\sim 50\%$. In this stellar mass range, halos in \simba\ become less multiphase in nature and more homogeneous, with a growing proportion of mass and metals in the hot phase of the CGM, largely due to stronger AGN feedback at the high mass end \citep{appleby_2021}. The lower CDDF of high mass galaxies compared with the overall CDDF demonstrates that this leads to a reduction in number of these CGM absorbers, as CGM gas is heated into higher ionisation states such as \ion{O}{vii} that are not traced in the ultraviolet~\citep[see][]{bradley_2022}.

In summary, \HI\ is the most abundant species in the CGM of \simba\ galaxies, with a large absorber population out to the high column density end. Star formation activity and stellar mass of the central galaxy both play a role in the absorber statistics of the CGM. The CDDF for each species is lower for quenched galaxies than star-forming galaxies, while the CDDFs for green valley galaxies is more similar to quenched than star-forming galaxies, suggesting that the effects of quenching are apparent in the CGM before star formation fully halts in the central galaxy. High mass galaxies also have lower absorber CDDFs due to the growing proportion of the CGM that is made up of hot gas. Numerical convergence tests conducted in our previous work indicate that increased particle resolution results in a moderate increase in equivalent width, particularly for \HI\ and \OVI\ absorption in the outskirts of halos, suggesting that these CDDFs (produced from the fiducial \simba\ volume) represent minimum CGM absorber counts \citep{appleby_2021}, particularly at the low column density end. In addition, as can be seen in Figure \ref{fig:compare_deltaN} of Appendix \ref{sec:appendix}, the Voigt profile fitting routine can overestimate \HI column densities in Lyman Limit Systems due to the saturated nature of the absorption. As such, the CDDFs presented here for \HI\ are likely to be overestimated at the high column density end.

\subsection{Physical conditions of absorber sample}\label{sec:histograms}

\begin{figure*}
	\includegraphics[width=\textwidth]{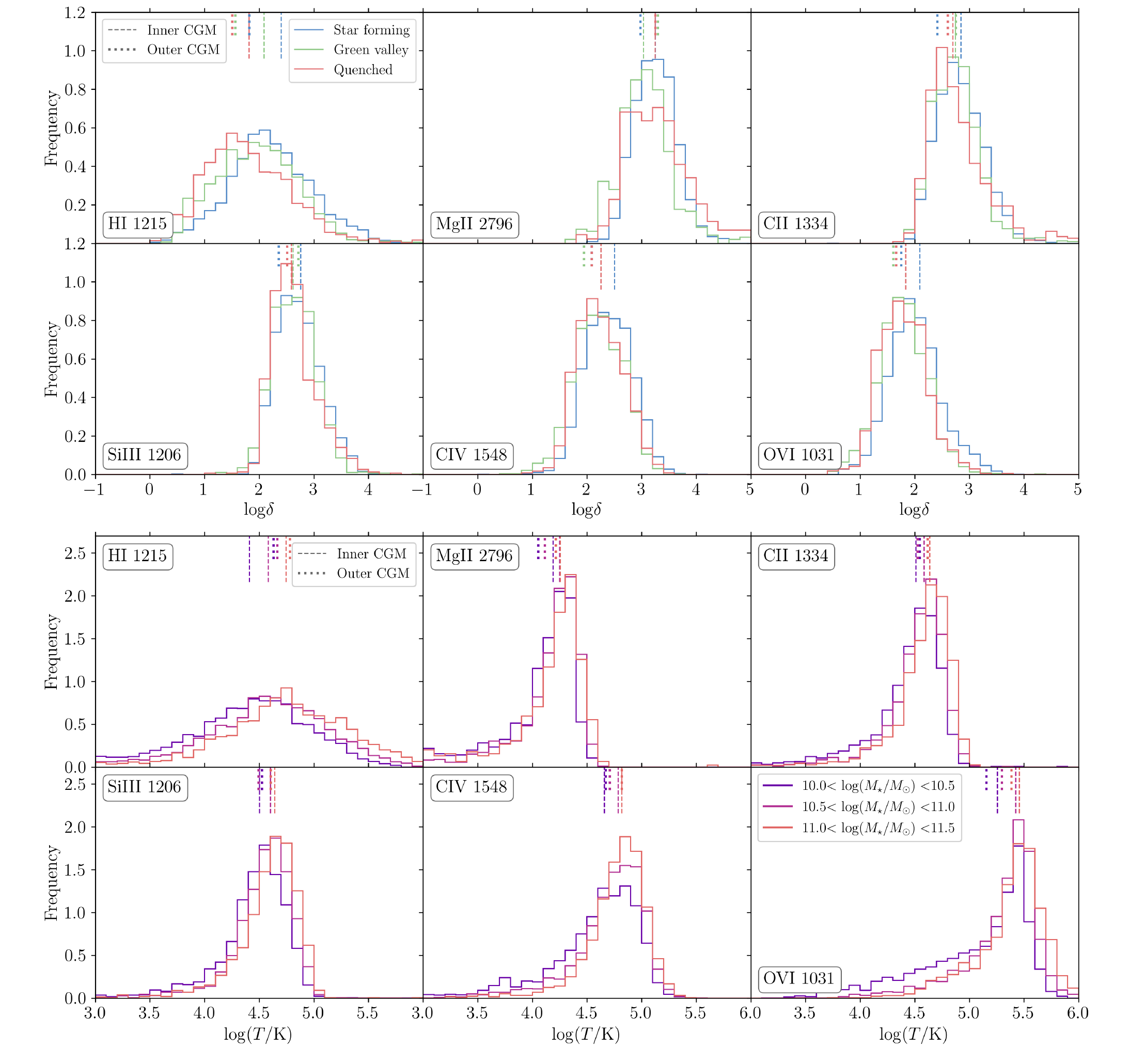}
	\vskip-0.1in
    \caption{Histograms of overdensity (top) and temperature (bottom) at the fitted positions of CGM  absorbers for each of the ions we consider. Overdensity histograms are split by the type of central galaxy: star forming (blue), green valley (green) and quenched (red). Temperature histograms are split by the stellar mass of the central galaxy: blue (low), purple (intermediate), and orange (high). Vertical lines in the upper portion of each panel indicate the medians for each galaxy sample when split by inner (dashed) and outer (dotted) CGM LOS.} 
    \label{fig:absorber_deltaT_hist}
\end{figure*}

We now turn to the underlying physical conditions that give rise to \HI\ and metal absorption in the CGM by examining the distribution of CGM absorbers in overdensity-temperature phase space. We obtain underlying physical conditions for each absorber by selecting column density-weighted properties (computed during the spectral generation process, see \S\ref{sec:spectra}) at the pixel nearest to the absorber's fitted LOS wavelength. Thus the results presented here represent the gas conditions that are contributing most to the absorption. Figure~\ref{fig:absorber_deltaT_hist} shows histograms of overdensity (upper panels) and temperature (lower panels) of the absorbers for the six ions we consider. Overdensity is defined as $\delta = \rho_m/\Bar{\rho}_m$, where $\rho_m$ is the matter density, and $\Bar{\rho}_m$ is the mean matter density of the universe. Each histogram is normalised such that the area sums to unity. Overdensity histograms are split into absorbers from star forming (blue), green valley (green) and quenched (red) galaxies. Temperature histograms are split by the stellar mass of the central galaxy: blue (low), purple (intermediate), and orange (high). In addition, vertical lines at the top of each panel indicate the medians of each galaxy sub-sample, when split into inner ($r_\perp/r_{200} = 0.25$ and $0.5$ exactly;  dashed lines) and outer ($r_\perp/r_{200} = 0.75, 1\ {\rm and}\ 1.25$ exactly;  shorter dotted lines) CGM LOS, following the inner and outer CGM definitions of \cite{oppenheimer_2018b}. We include lines at $r_\perp/r_{200} = 1.25$ to reflect the fact that the outer boundary of the CGM is not well-defined. 

The \HI\ absorbers trace gas across a broad range of physical conditions, both in terms of overdensity and temperature. Neutral \HI\ absorption can occur in a wide range of conditions due to the high oscillator strength of the \HI\ transition, though it traces hotter gas less well owing to collisional ionisation effects. Additionally, the presence of \HI\ does not rely on the metal enrichment of the gas, thus less enriched lower-density regions farther from the galaxy can still yield strong absorption.

By comparison, the overdensity and temperature distributions for the metal lines are generally narrower, because individual metal ions tend to trace specific phase space conditions. The median overdensity for metal line absorbers moves to lower overdensities and higher temperatures for the higher ionisation species; we will examine this relationship with ionisation potential in more detail in section \ref{sec:typical}.  This is expected since atoms tend to lie in lower ionisation states for denser, cooler gas.   Additionally, chemical enrichment tends to be strongest close to the galaxy where the CGM is densest, yielding more favorable conditions for low ion absorption.  Hence e.g. \MgII\ peaks at $\delta\sim 10^3$, while \OVI\ peaks at $\delta\sim 10^2$, similar to \HI.  Meanwhile, the temperature histograms show low ions peaking at $T\la 10^{4.5}$K, while \OVI\ is most commonly found near its collisional ionisation peak temperature of $10^{5.5}$K, albeit with a significant tail to lower temperatures representing photoionised gas. We will examine the relative contribution to the metal absorbers from collisional and photoionised gas in more detail in section \ref{subsec:ionisation}.

In the upper panels of Figure~\ref{fig:absorber_deltaT_hist} we separate the absorber sample by the type of central galaxy (star forming, green valley or quenched) and compare CGM absorber overdensity distributions. Absorbers in the CGM of star forming galaxies typically occur in gas that is more overdense than the absorbers around green valley and quenched galaxies, although the effect is small. For \HI\ absorbers in the inner CGM, the median overdensity occurs $\sim$0.3 and $\sim$0.6 dex higher for star forming galaxies than green valley and quenched galaxies, respectively. For absorbers in the outer CGM, the median overdensity is similarly higher around star forming than green valley galaxies, but there is little difference between green valley and quenched galaxies.  

For each ion, the green valley absorber distribution more closely resembles that of the quenched galaxies than the star forming galaxies. This supports our interpretation of the CDDF that the CGM of green valley galaxies shows signs of being quenched before star formation completely ends in the central galaxy. Although we do not show it here, the overdensity distributions are largely independent of galaxy stellar mass. 

For the temperature distributions in the lower panels of Figure~\ref{fig:absorber_deltaT_hist}, we separate the absorber sample by the stellar mass of the central galaxy, since it turns out that the temperature distributions are mostly independent of star formation activity.  The peak of the temperature distribution clearly increases with stellar mass, since galaxies of higher stellar mass typically live in higher mass halos with higher virial temperatures. Galaxies in the intermediate and high stellar mass bin produce absorbers which broadly follow the same temperature distributions. In contrast, galaxies in the lowest stellar mass bin have proportionally more absorbers in the low temperature tail of the distribution.

The median temperatures for these distributions are indicated by ticks along the top axis separated into inner and outer CGM, with the long ticks representing the former and the short ones the latter.
Absorbers from the inner CGM trace higher overdensities than those from the outer CGM. This trend is clearest for the \HI, \CIV\  and \OVI\ absorbers. The trend with impact parameter is not present for the \MgII, which was also the case with \cite{ford_2013}. We have seen from the \MgII\ CDDF that there are relatively few \MgII\ absorbers. Similarly, there is a slight trend with temperature for the \HI\ absorbers, with absorption in the inner CGM tracing preferentially cooler gas. For the metal lines there is very little difference between the median temperatures for inner and outer CGM.

To summarise, \HI\ traces a wide range of physical conditions in terms of overdensity and temperature, while the distributions for the metal absorbers are narrower, with peaks that decrease in overdensity and increase in temperature with increasing excitation energy. Absorbers from the inner CGM trace denser gas than those from the outer CGM; in contrast there is little difference in the absorber temperatures between the inner and outer CGM, except in the case of \HI. Separating the absorbers by the star formation activity of their central galaxies demonstrates that CGM absorbers from star forming galaxies typically arise from denser gas than those from green valley or quenched galaxies. The absorber temperature is largely set by the stellar mass of the central galaxy.  While these trends are subtle, they suggest that the ensemble of CGM absorption lines hold some information about the phase of CGM gas.

\subsection{CGM absorbers in cosmic phase space}\label{sec:phase_space}

\begin{figure*}
    \centering
    \begin{subfigure}[b]{\textwidth}
        \includegraphics[width=1\linewidth]{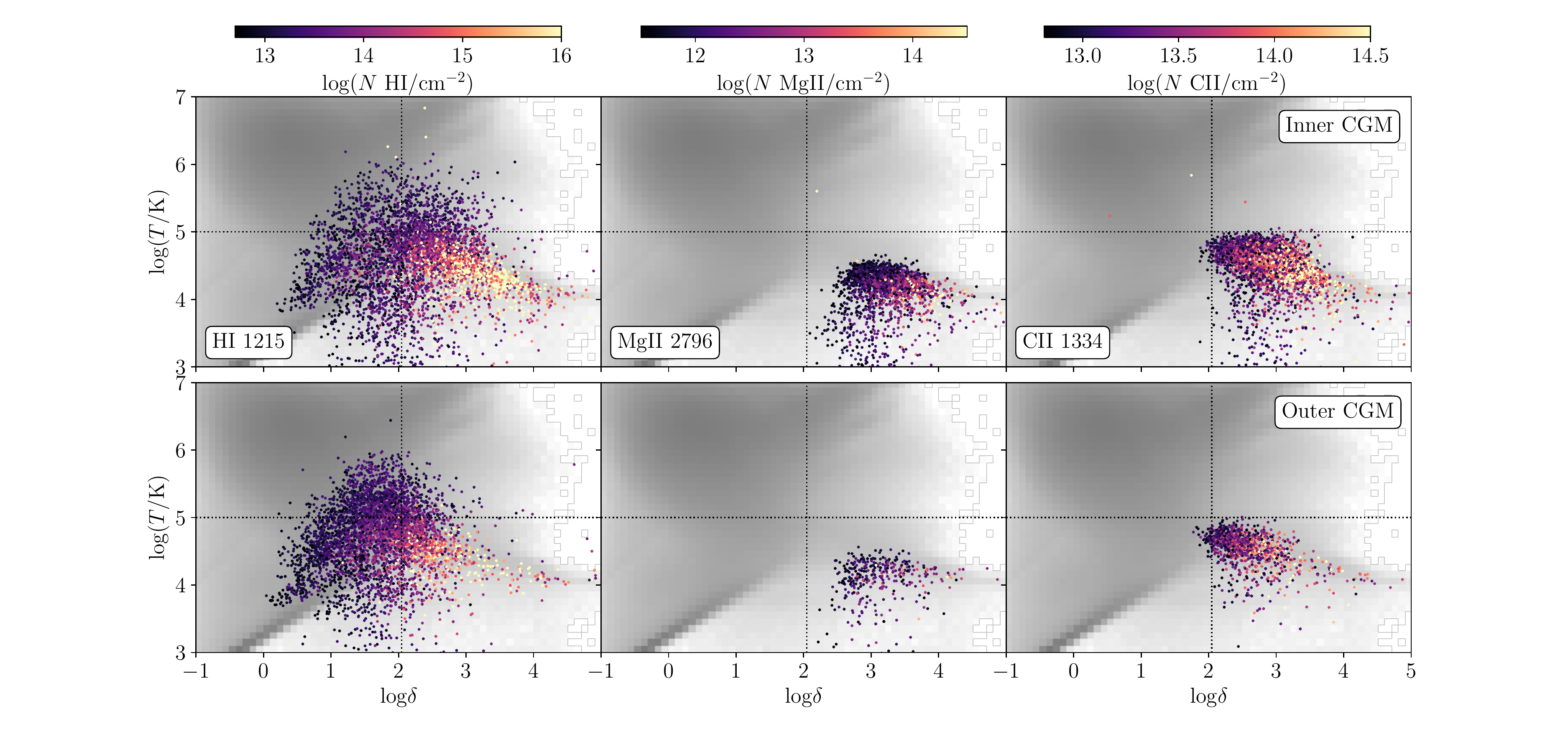}
    \end{subfigure}
	
	\begin{subfigure}[b]{\textwidth}
        \includegraphics[width=1\linewidth]{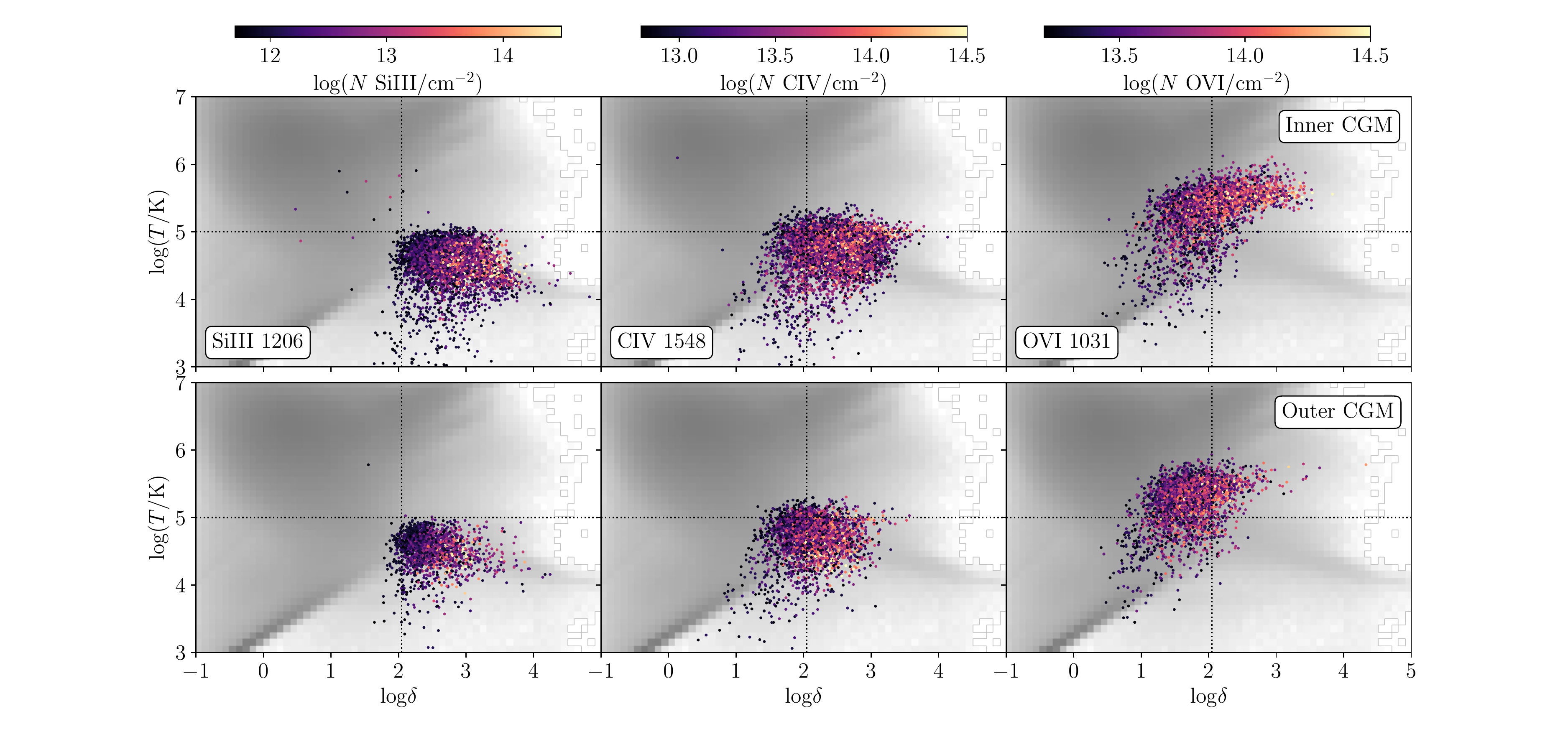}
    \end{subfigure}
	
	\vskip-0.1in
    \caption{Gas temperature against overdensity at the fitted position of CGM absorbers for the 6 species we consider, colour-coded by absorber column density. In each plot the top and bottom panels show absorbers in the inner and outer CGM, respectively. The upper plots show \HI\ (left), \MgII\ (middle) and \CII\ (right) absorbers, while the lower plots show \SiIII\ (left), \CIV\ (middle) and \OVI\ (right). The lower colour scale limits are the completeness limits and are different for each ion. The background greyscale in each panel shows the global phase space distribution for all gas along all LOS (this is identical for each species).}
    \label{fig:absorber_deltaTN}
\end{figure*}

We now combine the overdensity and temperature information into 2D phase space plots to gain a more complete picture of the absorber properties and examine their relationship with absorption strength. Figure \ref{fig:absorber_deltaTN} shows the temperature-overdensity phase space for the CGM absorbers, colour-coded by the absorber column density. Upper plots show \HI\ (left), \MgII\ (middle) and \CII\ (right) absorbers, while the lower plots show \SiIII\ (left), \CIV\ (middle) and \OVI\ (right). In each case the upper panels show absorbers in the inner CGM ($r_\perp/r_{200}$ of 0.25 and 0.5) and the lower panels show those in the outer CGM ($r_\perp/r_{200}$ of 0.75, 1, and 1.25). The lower limits of the column density colour scale are the completeness limits for each ion. In this plot we show all absorbers above the completeness limits, making no distinction between the different types of central galaxy. The phase space distribution for all gas along all the LOS is shown in greyscale. Compared with the phase space distribution for the entire original simulation at $z=0$ \citep[see][]{christiansen_2020}, the distribution from LOS gas is biased towards higher densities as they are selected to probe in and around the CGM of galaxies.

% Inner vs outer CGM
It is clear by comparing the upper and lower panels of Figure \ref{fig:absorber_deltaTN} that there are more identified absorbers for each metal ion in the inner CGM than in the outer CGM, and that high column density absorbers in particular are rarer in the outer CGM. The difference in number of absorbers is smaller for \HI, however the outer CGM still has preferentially fewer strong \HI\ absorbers. This is consistent with previous work examining halos in the \simba\ simulation that demonstrate a decreasing radial profile for density \citep{sorini_2020, sorini_2021} and EW \citep{appleby_2021}. The similarity between the phase space distributions in the inner and outer CGM illustrates that these absorbers arise from broadly the same physical conditions, but that such conditions are rarer in the outer CGM. This is particularly true for \MgII\ absorbers, which are almost non-existent in the outer CGM. 

% Cosmic phases
Following \cite{dave_2010}, the vertical lines in Figure \ref{fig:absorber_deltaTN} represent the approximate overdensity threshold at the boundary of a virialised halo, based on \citet{kitayama_1996}:
\begin{equation}
\label{eq:dth}
    \delta_{\rm th} = 6\pi^2(1 + 0.4093(1/f_\Omega -1)^{0.9052})
\end{equation}
where $f_\Omega$ is:
\begin{equation}
    f_\Omega = \frac{\Omega_m(1+z)^3}{\Omega_m(1+z)^3 + (1-\Omega_m - \Omega_\Lambda)(1+z)^2 + \Omega_\Lambda}
\end{equation}
At $z=0$, the overdensity threshold is $\delta_{\rm th} \sim 111$ for our cosmology. The horizontal lines represent the temperature threshold $T_{\rm th} = 10^5$K, which is the conventional definition for the Warm-Hot Intergalactic Medium (WHIM) gas \citep{cen_1999}; temperatures above this threshold cannot easily be reached without gravitational shock heating or feedback. \cite{dave_2010} use these thresholds to define four regions of cosmic phase space: {\it condensed} gas which is cool and dense; {\it hot halo} gas which is dense and shock heated; {\it diffuse} gas which is cool and mostly photoionised; and the {\it WHIM}, which is diffuse and hot. Only gas in the condensed or hot halo phases can be considered part of the CGM, as less dense gas is typically not bound to any halo. It was explicitly verified by \cite{sorini_2021} that the cumulative mass fraction of the diffuse and WHIM phases identified based on the overdensity threshold given by equation~\eqref{eq:dth} is very close to the value obtained by summing over the mass of gas elements outside the boundaries of haloes in \simba.

From Figure \ref{fig:absorber_deltaT_hist} we have seen that \HI\ absorbers span a wide range of physical conditions; Figure \ref{fig:absorber_deltaTN} shows that these absorbers are spread across the four categories of phase space, with the high column density absorbers ($N_{\rm HI} > 10^{15}{\rm cm}^{-2}$) absorbers arising from cool, condensed gas. Absorbers from the hot halo, WHIM and diffuse gas phases are typically weak, close to or below the completeness limit.

Absorption from the low ionisation metal lines (\MgII, \CII, and \SiIII) lies almost exclusively in the condensed phase; the temperature of these absorbers rarely rises above $T_{\rm th}$.The highest column density \CII\ absorbers lie along the same locus in the condensed phase as the strong \HI\ absorbers; this is not apparent for \MgII and \SiIII as these metal absorbers are weaker. For \CIV, these absorbers lie mainly across the condensed and diffuse regions of phase space, with the bulk of absorbers arising from condensed gas. In contrast to the lower ions, the temperatures of \CIV\ absorbers can rise above $T_{\rm th}$, and there is a contingent of these absorbers arising from both the WHIM and hot halo phases. There is however little trend in the column density of \CIV\ absorbers; the strongest absorbers are only mildly biased towards higher densities. Finally in the case of \OVI, these absorbers exist across the diffuse, WHIM and hot halo regions, with the highest column density absorbers arising from hot halo gas. In contrast to the other ions, \OVI\ absorbers rarely arise from cool, condensed gas. The phase space distributions of \MgII, \CII, \CIV\ and \OVI\ CGM absorbers presented here is consistent with measurements of halo metals in the Illustris-TNG simulations \citep{artale_2022}.

\begin{figure}
	\includegraphics[width=\columnwidth]{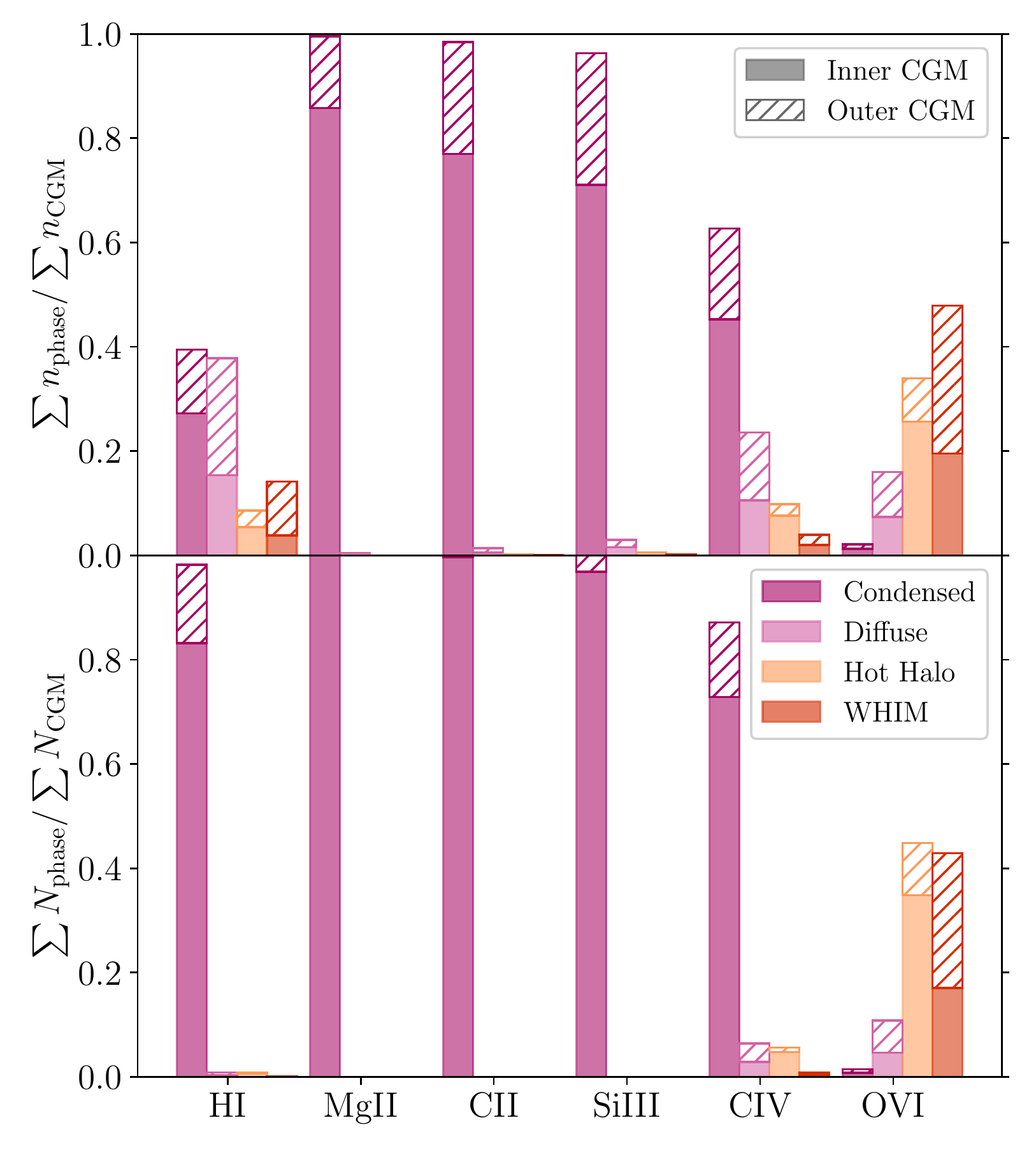}
	\vskip-0.1in
    \caption{{\it Top:} the number fraction of identified absorbers in each region of cosmic phase space, for each of the species we consider. Fractions are split into inner (solid) and outer (hatched) LOS. {\it Bottom:} as above, showing the fraction of total column density.} 
    \label{fig:phase_bar}
\end{figure}

Figure \ref{fig:phase_bar} quantifies these trends, showing the fraction of absorption arising from each region of cosmic phase space for each ion. This is computed as the fraction of the total number of fitted line profiles in each phase (top panel) and as the fraction of total column density in each phase (bottom). The fractions are split into the inner and outer CGM, represented as solid and hatched bars, respectively; the inner CGM clearly dominates the absorber samples. The number fractions show the intuitive expected fractions from examining Figure \ref{fig:absorber_deltaTN}, whereas the absorption fractions demonstrate how the particular regions of phase space would dominate the actual observed absorption. The total column density of the LOS is a robust measurement of the overall absorption, as we have verified that the total LOS equivalent width from profile fitting largely corresponds to the equivalent width from direct measurement. Hence the total absorption can be comfortably compared between works with different fitting procedures. However the number fractions is less comparable since arguably the precise number of absorbers is dependent on the details of the profile fitting procedure.

The number fractions of \HI\ absorbers shows that 77\% arise from cool gas (39\% and 38\% in the condensed and diffuse phases, respectively), 9\% from hot halo gas and 14\% from the WHIM. In addition, the inner CGM represents 52\% of all \HI\ absorbers and 69\% of the condensed \HI\ absorbers. For the low metal ions, virtually all of the absorbers are condensed gas. These mostly arise from the inner CGM, with the fraction in the inner CGM decreasing with ionisation energy for these low metal ions. More generally, the fraction of absorbers in the condensed phase decreases with increasing ionisation energy. \CIV\ in contrast has a significant fraction of absorbers (37\%) not in condensed gas. For \OVI, the largest fraction of absorbers comes from the WHIM (48\%) while the hot halo and diffuse gas comprise 34\% and 16\% of the absorbers, respectively. The inner and outer CGM contribute roughly equally to the absorption from the WHIM (41\% and 59\%, respectively), whereas the hot halo absorption seen from the inner CGM is less common in the outer CGM.

Examining the fraction with respect to total absorption demonstrates that while our absorber sample represents a wide range of overdensities and temperatures, the majority of observed absorption arises from inner CGM gas in the condensed phase, except in the case of \OVI. \HI\ absorption is split into 83\% condensed in the inner CGM and 15\% condensed in the outer CGM, with only 2\% arising from the remaining phases. For the low ions \MgII, \CII\ and \SiIII\ the condensed gas fraction from the inner CGM is essentially 100\% of the total absorption (97\% for \SiIII). For \CIV, although 37\% of absorbers lie in the diffuse, hot halo and WHIM phases, these collectively contribute only 13\% of the total \CIV\ absorption. Finally for \OVI\ the proportion in the hot halo phase rises to 45\% when considering total absorption. This increase comes at the expense of all other phases, such that the WHIM and hot halo gas contribute equally to the majority of the absorption (88\%). Since gas in the WHIM phase as defined by \citet{dave_2010} is not strictly part of the CGM, we conclude that only about half ($\sim46\%$) of the observed \OVI\ absorption actually arises from CGM gas.

In summary, there are fewer absorbers in the outer CGM compared with the inner CGM, regardless of absorber species. \HI\ (and \CIV) absorbers are distributed across all regions of phase space, but most of the overall absorption (98\% and 87\%) arises from condensed gas. Absorption from low ions (\MgII, \CII, \SiIII) arises almost exclusively from condensed gas, both when considered in terms of number of absorbers and total overall absorption. In contrast, \OVI absorbers arise predominantly from gas in the WHIM or hot gas phase, with less than half of absorbers arising from gas that is considered part of the CGM.

\subsection{Sources of ionisation}
\label{subsec:ionisation}

Earlier work using \simba\ to study the CGM has found that the choice of photoionising background makes a considerable difference to such measures of the CGM absorption structure \citep{appleby_2021}. Here we extend this analysis by examining the relative contributions of collisional ionisation versus photoionisation to the ionisation structure of the CGM. We do so for \CIV\ and \OVI\ as we have seen in \S\ref{sec:phase_space} that these ions have significant fractions of absorbing gas above the canonical temperature limit for photoionisation ($T_{\rm photo}$). Since the photoionising sources are unable to ionise above $\sim 10^5$K, we expect that any absorbers above $T_{\rm photo}$ must arise from collisional ionisation; hence the hot gas absorption fraction is a reasonable proxy for the contribution to the total column density from collisional sources. Previous simulations have shown that high excitation ions trace collisionally ionised gas at high temperatures \citep{ford_2013}. We adopt temperature thresholds of  ${\rm log} (T_{\rm photo}/ {\rm K}) = 4.8$ for \CIV\ \citep{mauerhofer_2021} and ${\rm log} (T_{\rm photo}/ {\rm  K}) = 5$ for \OVI\ \citep{bohringer_1998, oppenheimer_2009a}. 

\begin{figure}
	\includegraphics[width=\columnwidth]{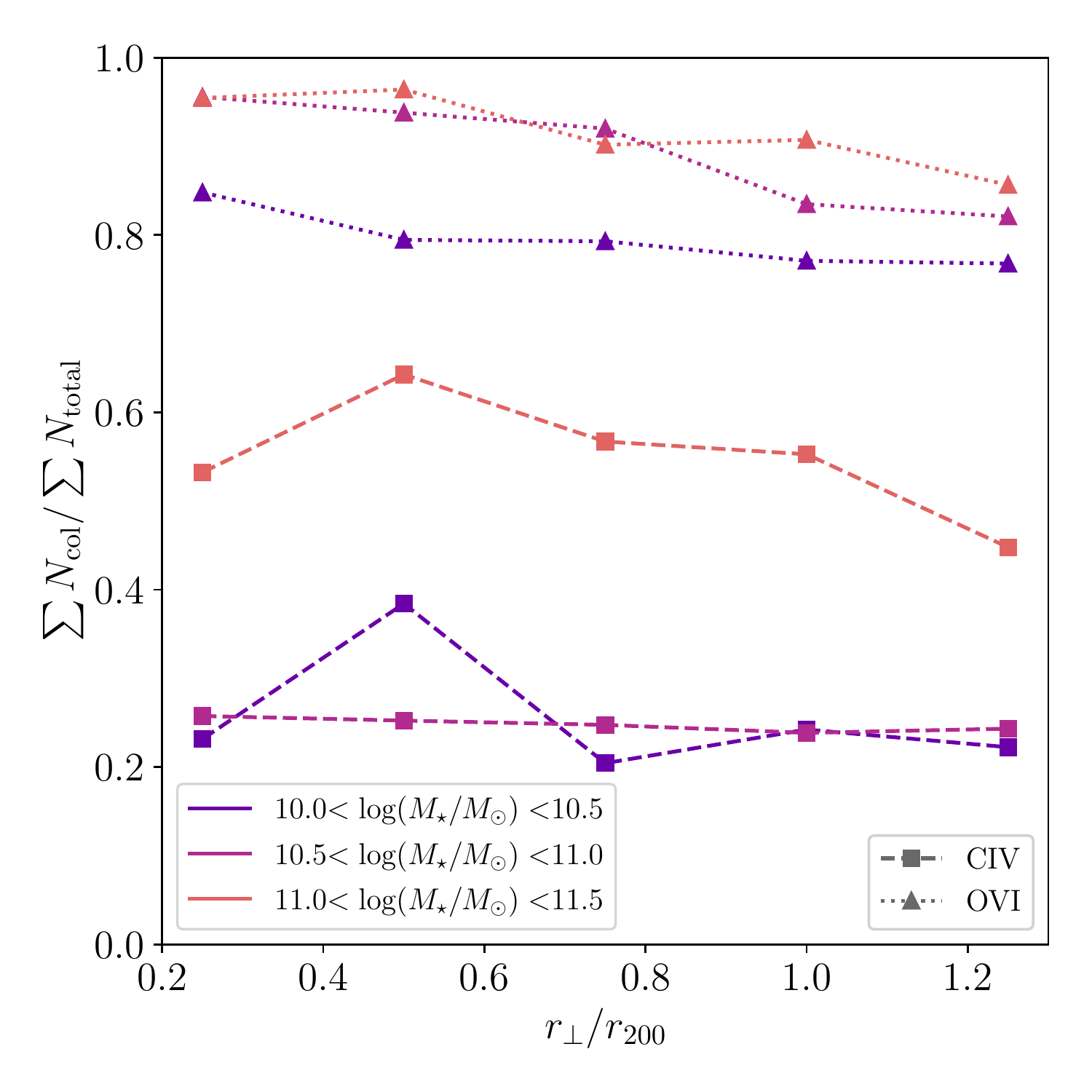}
	\vskip-0.1in
    \caption{The fraction of absorption arising from collisionally ionised gas as a function of impact parameter, for \CIV\ (dashed lines with squares) and \OVI\ (dotted lines with triangles). Line colours represent broad galaxy stellar mass bins.} 
    \label{fig:fcol}
\end{figure}

Figure \ref{fig:fcol} shows the fraction of total column density arising from hot gas above $T_{\rm photo}$ as a function of impact parameter, for \CIV\ (dashed with squares) and \OVI (dotted with triangles). We separate the absorbers by the stellar mass of their central galaxy: low (blue), intermediate (purple) and high (orange). Both ions show significant fractions of absorbers in collisionally ionised gas. \OVI\ absorbers are at a minimum $\sim 80\%$ collisionally ionised, consistent with high \OVI\ collisional fractions measured for this mass regime in the NIHAO simulations \citep{gutcke_2017}. Similarly, \cite{oppenheimer_2009a} found that while much of the IGM \OVI\ absorption is photoionised, the strongest \OVI\ absorbers (i.e. those who contribute most to the total \OVI\ column density) are likely collisionally ionised and trace recent enrichment within galaxy halos. For \CIV, the minimum collisional fraction is $\sim 20\%$. Large collisional fractions for the high excitation ions naturally follows from our discussion of the high proportion of \OVI\ absorbers which arise from hot halo gas or the WHIM. These results demonstrate that collisional ionisation is the dominant ionising source for these intermediate/high excitation ions.

We have tested the impact of the ionising background on our results by re-generating the \CIV and \OVI spectra using the \citet{haardt_2012} photoionising background instead of the \citet{faucher-giguere_2020} background. The results are mostly unchanged - we find that the absorbers occupy the same regions of phase space and that the fraction of \OVI\ from collisional sources is qualitatively the same. However for \CIV\ there is a slight ($\sim$5\%) increase in the fraction of absorption from collisions. This arises since the \citet{haardt_2012} background predicts lower fractions of \CIV in the photoionised temperature regime, at the level of $\sim$0.15 dex at the photoionisation limit of $10^{4.8}{\rm K}$ and $\delta_{\rm} = 10^2$ (the precise decrease is sensitive to temperature and overdensity). This leads to (slightly) reduced \CIV absorption from the ionising background and thus a higher contribution from collisional sources.

We break our results down further as a function of stellar mass and impact parameter. The contribution from collisions increases with stellar mass for both ions, which is a reflection of the increasing halo virial temperature for high mass galaxies (as discussed in \S\ref{sec:histograms}). For the \CIV\ absorbers there is little increase between the low and intermediate mass bins, but a substantial increase from of $\sim 30\%$ in collisions between the intermediate and high mass bins. Galaxies in the intermediate mass bin ($10^{10.5}\msolar < \mstar < 10^{11}\msolar$) have mass-weighted average CGM temperatures down to $10^{5.5}$K (see Figure \ref{fig:galaxy_sample}) and their gas-phase metals roughly evenly distributed across the ISM and cool, warm and hot CGM gas phases \citep{appleby_2021}. Thus gas below the photoionisation limit for \CIV\ is more common in low and intermediate mass halos, whereas the CGM of high mass galaxies ($\mstar > 10^{11}\msolar$) have a mass-weighted average CGM temperature of no less than $10^6$K and are more dominated by hot CGM gas (cit.), resulting in a large increase in collisional fraction. For the \OVI\ absorbers the largest increase in collisional fraction ($\sim 10\%$) occurs between the low and intermediate mass bins. In \S\ref{sec:histograms} we noted that the \OVI\ temperature distribution around low mass galaxies has a tail of low temperature absorbers; here we see that this photoionised tail contributes at most $\sim 20\%$ of the total absorption.

The collisional absorption fraction also slightly decreases with impact parameter, which reflects the downward radial trend in halo temperature as traced by X-rays in \simba\ \citep{robson_2020}. The radial trend is present for the \OVI\ absorbers in every mass bin; for \CIV\ absorbers the trend is strongest for the highest mass bin, while absorber fraction profiles for low and intermediate mass galaxies is broadly flat within the noise. The radial decrease is consistent with results from the NIHAO simulations, which show that photoionisation is more prominent on the outskirts of halos \citep{roca_fabrega_2019}. To summarise, the fraction of collisionally ionised absorption is $\sim 25-55\%$ for \CIV\ and $\sim 80-95\%$ for \OVI, increases with galaxy stellar mass, and decreases with impact parameter.

\subsection{Physical density versus column density}\label{sec:density}

Naively, one expects that higher density gas gives rise to higher column density absorption.  However, temperature and metallicity variations, along with geometrical effects, can complicate this relationship.
Here we take a closer look at the relationship between physical overdensity of the 3D gas elements and the measured 2D column densities of our absorber sample, and examine any dependence on the type of central galaxy. 

\begin{figure*}
	\includegraphics[width=\textwidth]{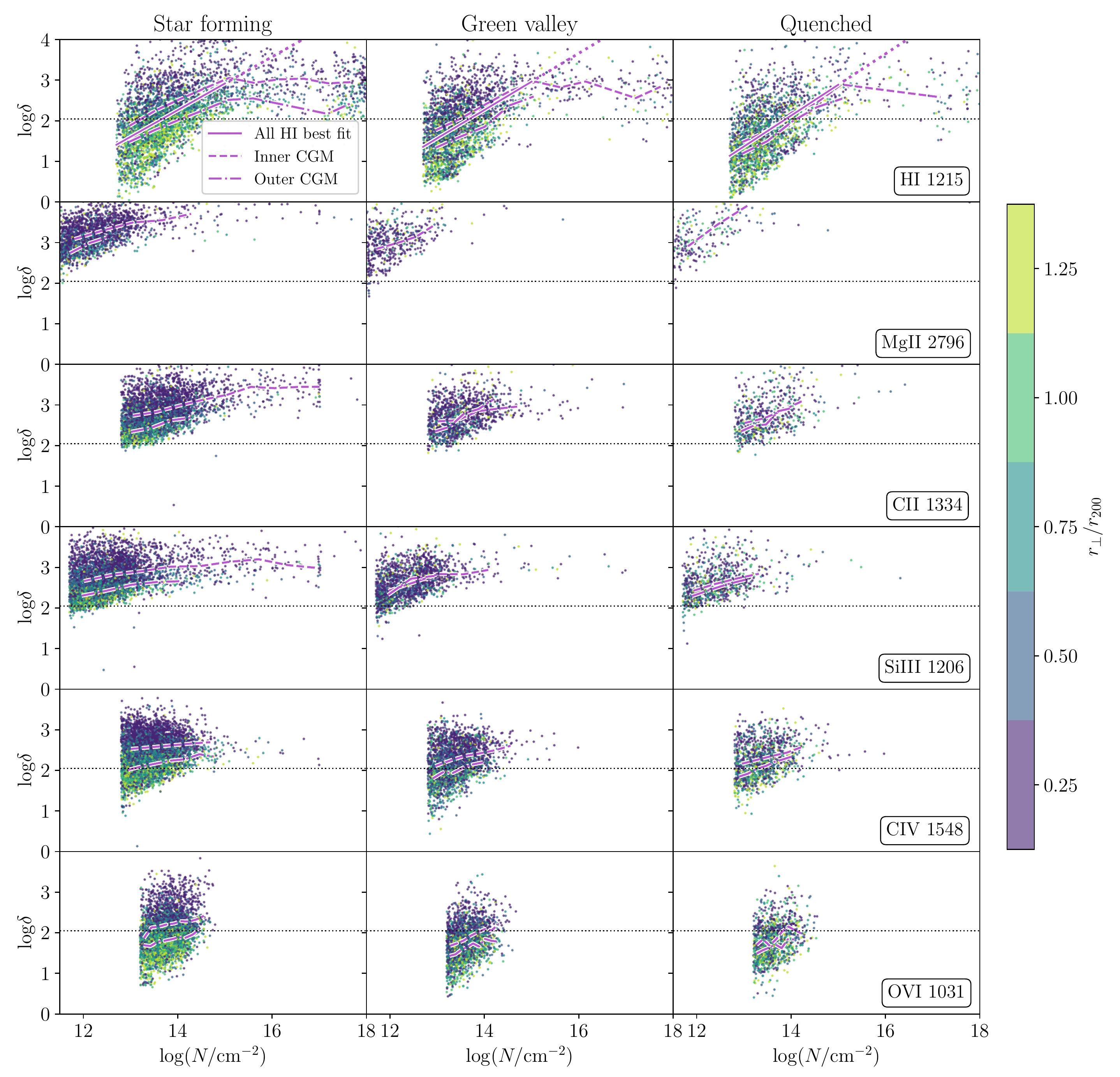}
	\vskip-0.1in
    \caption{Overdensity against column density for each of the species, split into star forming (left), green valley (middle) and quenched (right) galaxies. Points are colour-coded by the $r200$-normalised impact parameter. Purple dashed and dot-dashed lines show the running medians for galaxies in the inner and outer CGM, respectively. The solid purple lines indicate the best power law fit to the \HI\ absorber data between the completeness limit and $10^{15}$ cm$^{-2}$; the dotted purples lines show the best fit extrapolated to higher column densities. }
    \label{fig:Ndelta_r200}
\end{figure*}

Figure \ref{fig:Ndelta_r200} shows gas overdensity against absorber column density, colour-coded by $r_{200}$-normalised impact parameter. Results are split into absorbers arising around star forming (left), green valley (middle) and quenched (right) galaxies. The purple lines in each panel represent the running median for inner (dashed) and outer (dot-dashed) CGM absorbers; we show medians only where the column density bins contain at least 15 data points. The solid purple lines indicate the best fit to a power law for the \HI\ absorber data between the completeness limit and $10^{15}$ cm$^{-2}$; the dotted purple lines show the best fit extrapolated to higher column densities.

First we examine the overall results for the whole absorber sample. For each absorber species, there is a positive trend of increasing overdensity with column density, as broadly expected. For the metal lines, the cloud of absorbers moves to lower overdensities with increasing excitation energy. The running medians provide an estimate for the overdensity for observed absorbers at a given column density. Nonetheless, there is a significant scatter around this relationship.

Of the absorber species we include, \HI\ has the steepest increase of overdensity with column density. Between the completeness limit and $10^{15}$ cm$^{-2}$, the \HI\ absorbers follow a power law:
\begin{equation}
    {\rm log}_{10} \delta = a {\rm log}_{10} (N_{\rm HI} / {\rm cm}^{-2}) + b
\end{equation}
For star forming galaxies, the best fit parameters for the power law scaling are $a = 0.66$ and $b = -6.9$; for green valley galaxies these are $a = 0.7$ and $b = -7.5$; for quenched galaxies they are $a = 0.76$ and $b = -8.5$. These best fit parameters are reasonably similar to those found for \HI\ for IGM absorbers by \cite{dave_1999} ($a = 0.7$ and $b = -8.5$), despite the many differences between these simulations. The \HI\ absorbers in \simba\ trace higher overdensities for a given column density, perhaps simply owing to the higher densities found in the CGM compared with IGM gas, as well as changes in the assumed photo-ionising background. 

In addition, the CGM in Simba has a deficit of \HI\ absorbers at $N_{\rm HI} \sim 10^{17}$ cm$^{-2}$, and a cloud of absorbers at higher column densities ($N_{\rm HI} > 10^{17}$ cm$^{-2}$). Beyond $N_{\rm HI} > 10^{15}$ cm$^{-2}$ the overdensity no longer increases uniformly with column density, as the absorption features are likely saturated. These absorbers exist on the logarithmic part of the curve of growth and so it becomes difficult for the Voigt profile fitter to disentangle increases in column density vs. linewidth. The cloud of high column densities comes from gas at all impact parameters, but preferentially comes from the inner CGM. 

When we split the absorbers by the star formation rate of central galaxy, we see subtle differences in these absorber samples, as hinted by the star formation dependence of the power law best fit parameters. When compared with star forming galaxies, quenched galaxies have fewer absorbers overall and preferentially fewer at the highest overdensities and at the highest column densities, due to the heating of the galaxy halo via AGN feedback. The reduction in the number of absorbers around quenched galaxies is largest for low metal absorbers since AGN feedback heats gas out of the optimal cool, dense regime for low metal absorption \citep{christiansen_2020}. In general, absorbers around star forming galaxies above $\delta_{\rm th}$ come from the inner CGM, whereas absorbers below this threshold come from the outer CGM. However for quenched galaxies, this clear distinction in overdensity between inner and outer CGM is not seen; this is clear from the running medians for the inner and outer CGM absorbers, which lie close together for quenched galaxies. These results suggest that the CGM around quenched galaxies is more uniform in nature.

Despite these differences, overall we find that absorbers from around galaxies of different star formation activities do not come from fundamentally different types of gas, and it is not possible to cleanly separate our sample into different galaxy types by making cuts in overdensity-column density space. Although we do not show it here, we have also produced the converse version of Figure \ref{fig:Ndelta_r200} (coloured by the star formation activity and split into inner and outer CGM), and similarly found that absorbers from different types of galaxies occupy similar regions of phase space. We have also examined the stellar mass dependence and found that the impact of increasing stellar mass largely resembles that of decreasing star formation rate -- fewer absorbers for each species, but little difference in the distribution of absorbers in density space. These plots demonstrate a complex, degenerate relationship between star formation, stellar mass, absorption strength, and distance from the galaxy. We find that while the overall number and strength of absorption reduces when galaxies are quenched, conditions necessary for absorption do not change, and thus there is little change in the distribution of absorbers in phase space.

\subsection{Typical physical conditions of absorbers}\label{sec:typical}

\begin{figure}
	\includegraphics[width=\columnwidth]{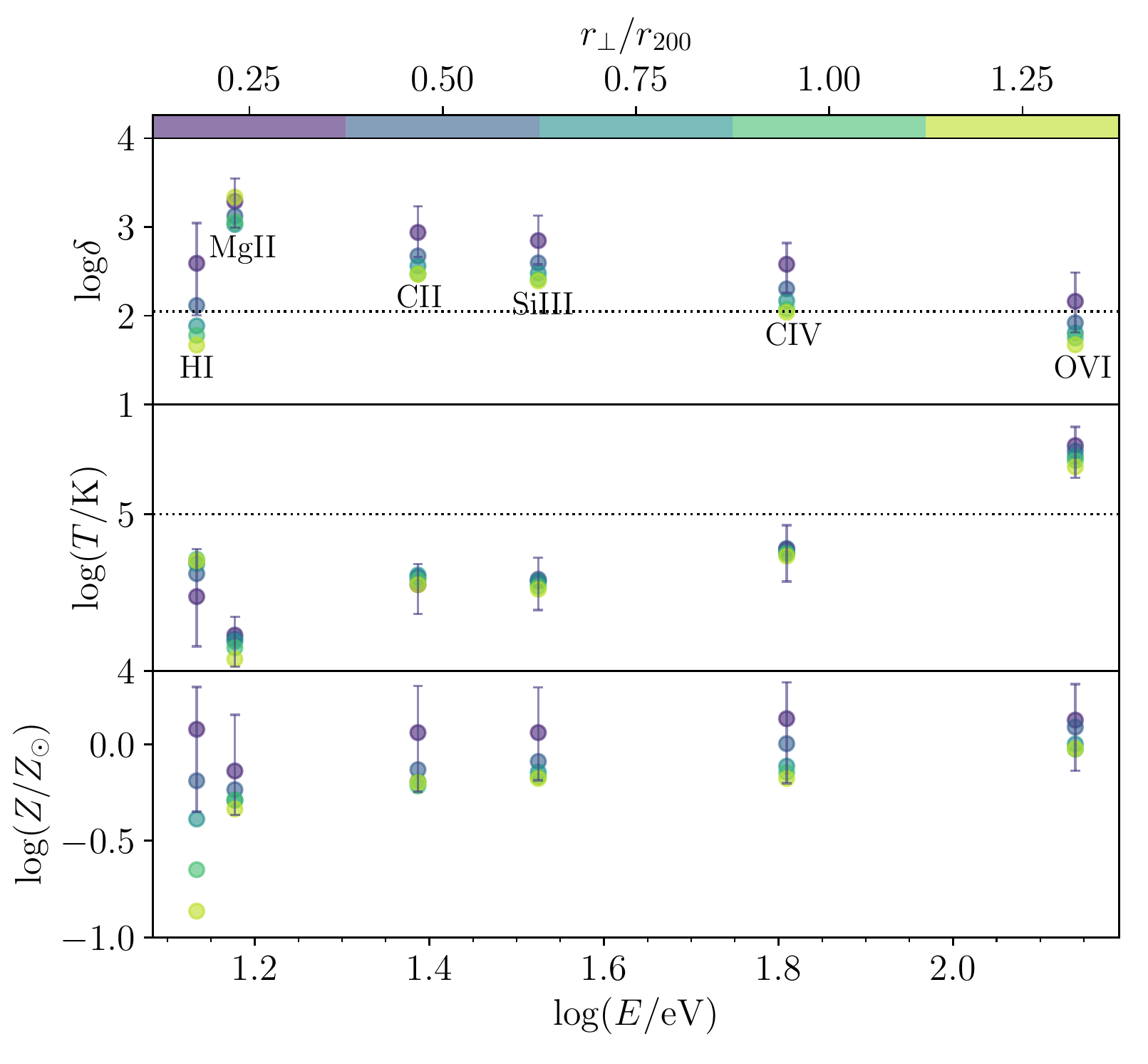}
	\vskip-0.1in
    \caption{Top panel: The column density-weighted median overdensity against ionisation energy for the 6 ions considered in this work. Points are colour-coded by the $r_{200}$-normalised impact parameter. Representative error bars are plotted for the $r_\perp/r_{200} = 0.25$ set of points, representing the 25-75 percentile range. Middle panel: as above, for column density-weighted gas temperature. Bottom panel: as above, for column density-weighted gas metallicity. Horizontal dotted lines in the top and middle panels represent the $\delta_{\rm th}$ and $T_{\rm th}$ thresholds, respectively. } 
    \label{fig:Nweighted}
\end{figure}

Having presented the physical gas conditions for our whole absorber sample, here we summarise these results by examining typical gas properties for each species. Figure \ref{fig:Nweighted} shows the column density-weighted median overdensity (top), temperature (middle) and metallicity (bottom) against ionisation energy for the six ions we consider.  For \HI, the metallicity is taken as the total metal mass fraction relative to solar abundance (0.0134), while for the individual ions it represents the solar-scaled abundance in the corresponding element. Column density-weighted medians represent the values at which 50\% of the absorption occurs above (or below) this value. Points are colour coded by the $r_{200}$-normalised impact parameter. We show representative error bars for the $r_\perp/r_{200} = 0.25$ set of points, representing the 25-75 percentile range. The horizontal lines in the top and middle panels indicate $\delta_{\rm th}$ and $T_{\rm th}$, respectively.

% HI
First we examine the typical conditions of the \HI\ absorbers. The typical overdensity of \HI\ absorbers decreases with increasing impact parameter, dropping below $\delta_{\rm th}$ at $0.75r_{200}$, indicating that absorbers seen at these large impact parameters are not typically part of the CGM. The typical temperature is ${\rm log} (T / {\rm K}) \sim 4.5-4.7$; there is a slight increase in temperature with increasing impact parameter, although in this case the increase is within the size of the interquartile spread. Lastly, the typical metallicity of \HI\ absorbers decreases with impact parameter; this reflects a decreasing metal enrichment of the CGM as a function of distance from the central galaxy.

% Metals
Median overdensity of metal absorbers decreases with ionisation potential and with increasing impact parameter, consistent with \cite{ford_2013} (except in the case of \MgII\ at high impact parameter; there are few \MgII\ absorbers at this distance from the central galaxy). At high impact parameters, the halo gas is more diffuse and thus metals are ionised preferentially by photoionisation, rather than via collisions -- this is reflected in the lower typical densities of metal absorbers at high impact parameter. For more detail on the relative contribution of the two sources of ionisation, see \S\ref{subsec:ionisation}. 

The median temperature of metal absorbers increases with ionisation potential. \MgII\ and \OVI\ have a weak dependence on impact parameter, with higher typical temperatures at lower impact parameters; the other metal ions have no dependence on impact parameter. \CII, \SiIII\ and \CIV\ require temperatures around $10^{4.5}$K, the canonical temperature threshold for photoionisation. \OVI\ absorption arises from warm gas that constitutes the volume-filling ambient gas of the CGM~\citep[e.g.][]{ford_2013,oppenheimer_2020a}.

The median metallicity of metal absorbers increases slowly with ionisation potential and decreases with impact parameter. As with \HI, this is a reflection of the underlying halo metallicity profile. However, the trend with impact parameter is much less strong than for \HI, because much of this absorption arises in dense gas close to galaxies, at least for lower ionisation absorbers.
At all impact parameters, the metal absorption arises from gas that is near the solar metallicity.  This is even true for \OVI\ at high impact parameters which arises predominantly in WHIM gas, potentially reflecting a bias for such absorption to arise in regions of relatively high enrichment even when the overall metallicity is lower~\citep{oppenheimer_2009a}.

\section{Redshift dependence}\label{sec:redshift}

Up until now, we have presented results for our sample of absorbers at $z=0$; here we examine the redshift dependence of our results using our higher redshift galaxy samples. We generate a set of synthetic absorption spectra for the galaxies in our $z=0.25, 0.5$ and 1 samples following an identical procedure to that for our $z=0$ sample (beyond $z=1$ these absorbers are redshifted out of the UV regime and as such different observational probes of the CGM are required). We fit Voigt profiles to these LOS as before to produce a sample of absorbers for each redshift.

\begin{figure*}
	\includegraphics[width=\textwidth]{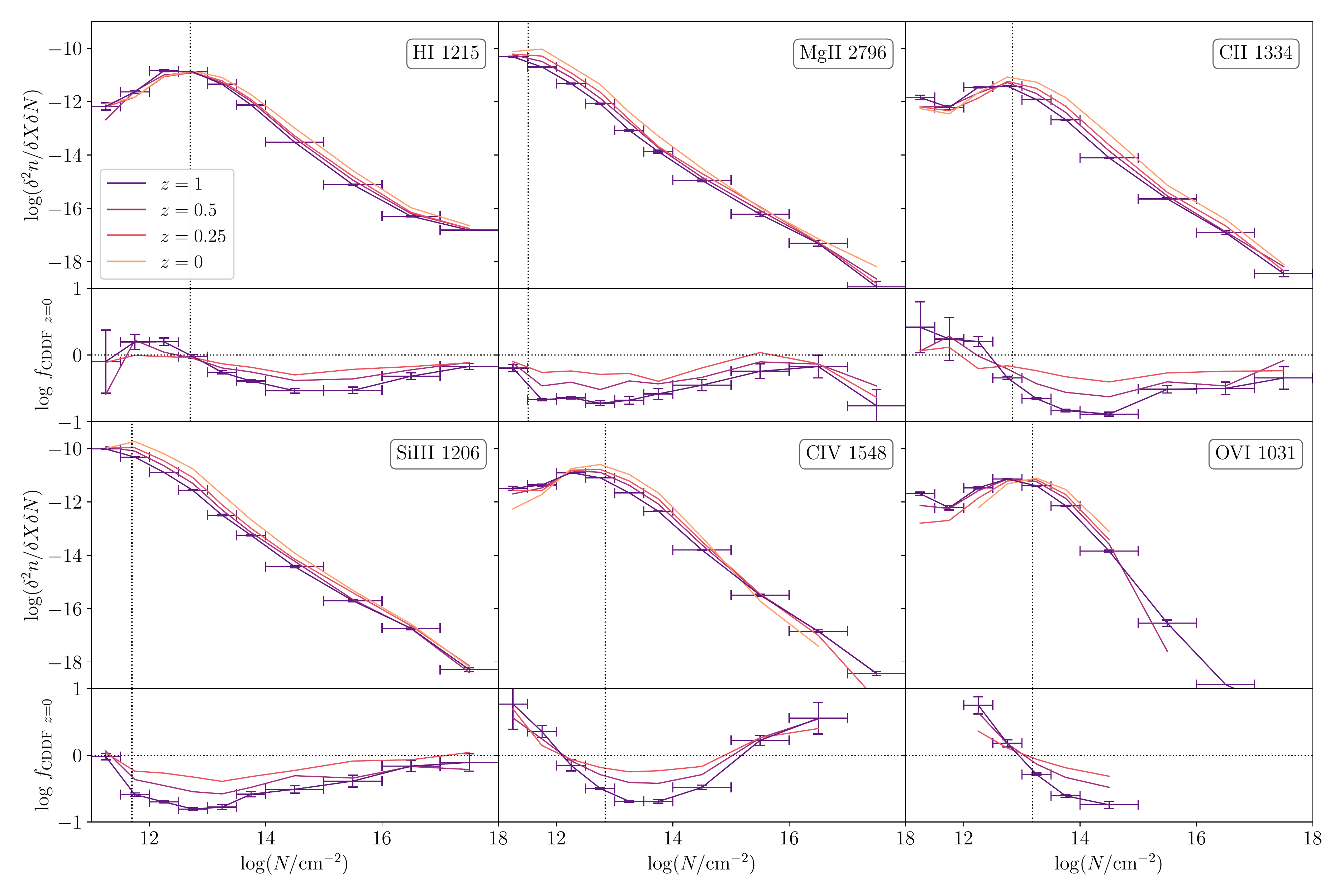}
	\vskip-0.1in
    \caption{Redshift dependence of the CDDF for each species considered in this work, shown at $z=1, 0.5, 0.25$ and $z=0$. The smaller panels show the fraction with respect to the CDDF at $z=0$. Vertical error bars on show representative cosmic variance uncertainties, computed using the $z=1$ sample; horizontal error bars indicate the width of the column density bins.} 
    \label{fig:cddf_redshift}
\end{figure*}

Figure \ref{fig:cddf_redshift} shows the redshift evolution of the CDDF for each species. The larger panels show the CDDFs, while the smaller panels show the CDDF fraction of the higher redshift samples with respect to the $z=0$ case. The horizontal dashed lines in the lower panels indicate the case where there is no redshift dependence. The vertical dashed lines indicate the $z=0$ completeness limits; the completeness limits for the higher redshift samples are lower than these, owing to an increase in number of absorbers with redshift. As such, above these column densities the CDDFs for each redshift can be considered representative of a complete sample. As in Figure \ref{fig:cddf}, the horizontal error bars show the width of the column density bins, where we have increased the bin width at the high column density end to account for low numbers. Vertical error bars show representative cosmic variance uncertainties using all galaxies and all LOS. These are computed as in \S\ref{sec:cddf}: the $x$ and $y$ sides of the simulation box are each split into quarters to form 16 cells across the face of the simulation; the absorbers are binned into the cells based on their $x-y$ positions; the cosmic variance uncertainties are computed by jackknife resampling across the 16 cells. 
For the CDDF fraction panels, the vertical error bars are computed by adding the $z=1$ errors in quadrature with the $z=0$ errors.

In general, for each species the normalisation of the CDDF at intermediate column density decreases with redshift. This does not represent a reduction in the number of absorbers within the simulation but mostly owes to the increase in the comoving path length $dX$; examining simple histograms of column density shows that the raw number and strength of absorbers increases with redshift. The increase is greatest for \HI, but is present for every species. At $z \geq0.5$, we find that \OVI\ absorbers exist outside of the narrow column density range seen at $z=0$. However, when we account for the cosmic evolution of the path length (which increases by a factor of 4.6 between $z=0$ and $z=1$), the CDDF decreases with redshift. At the high and low column ends of the CDDF, the large increase in the number of absorbers at $z=1$ compensates for the path length, and here we find that the difference with respect to the $z=0$ CDDF is reduced.

\begin{figure}
	\includegraphics[width=\columnwidth]{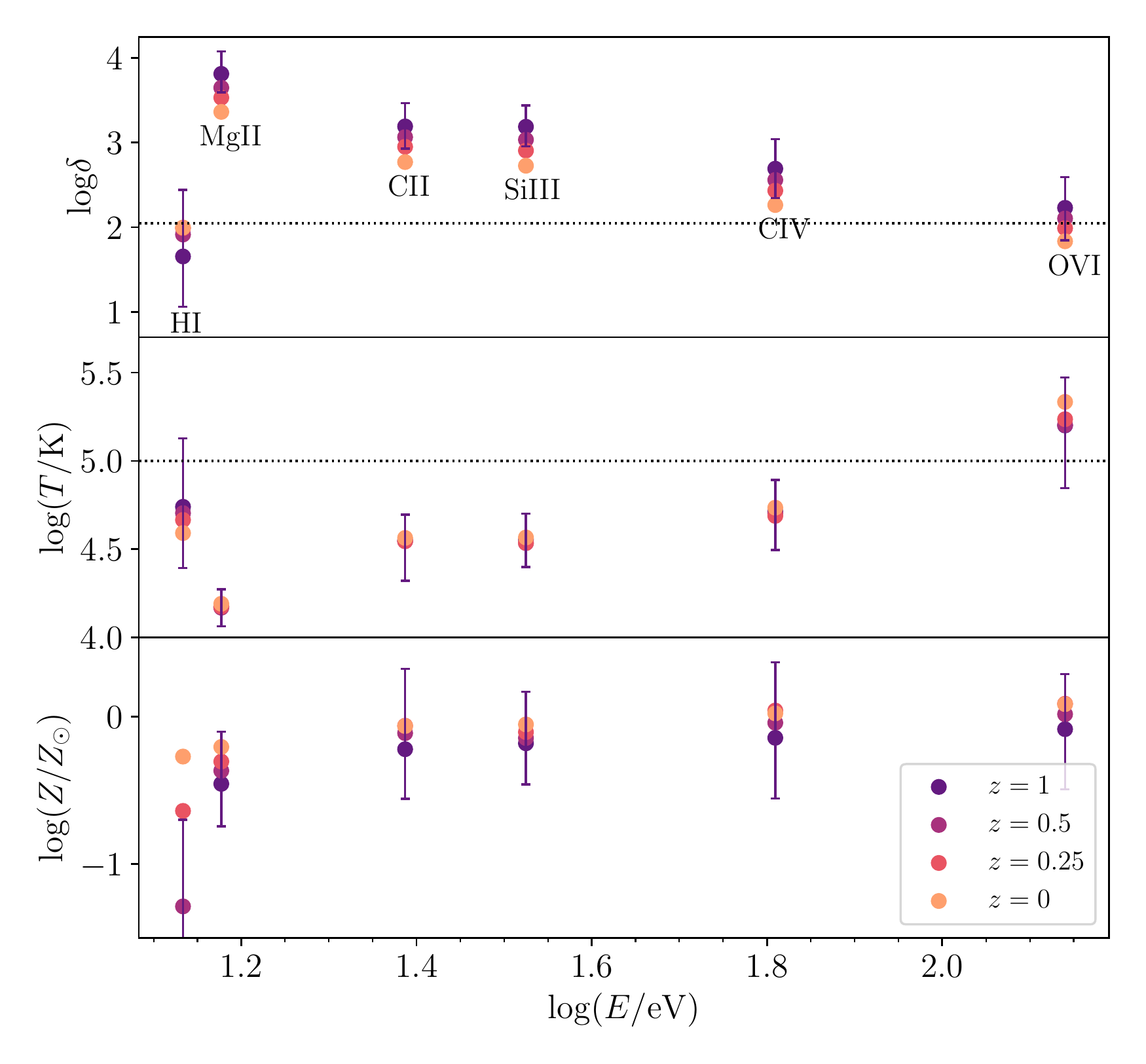}
	\vskip-0.1in
    \caption{As in Fig. \ref{fig:Nweighted}, showing the redshift dependence of the typical absorber conditions. The column density-weighted median absorber physical conditions for each species are plotted against ionisation energy: overdensity (top), temperature (middle) and metallicity (bottom). The different colours represent different redshifts. The horizontal lines in the top and middle panels indicate $\delta_{\rm th}$ and $T_{\rm th}$, respectively. The vertical error bars on the $z=1$ data points represent the 25-75 percentile range.} 
    \label{fig:Nweighted_redshift}
\end{figure}

Finally, we turn to the redshift dependence of the physical conditions of the CGM absorbers. Figure \ref{fig:Nweighted_redshift} shows the redshift evolution of the typical gas conditions for our absorber sample. As in Figure \ref{fig:Nweighted}, we plot the column density-weighted median overdensity (top), temperature (middle) and metallicity (bottom) against the ionisation energy of each species. Different coloured points represent the results for the different redshift samples. The vertical error bars represent the 25-75 percentile range for the $z=1$ sample. The horizontal dashed lines in the top and middle panels indicate $\delta_{\rm th}$ and $T_{\rm th}$, respectively.

For each metal ion, the column density-weighted median overdensity of the gas increases with redshift; at low redshift the halo gas is more diffuse due to the evacuating impact of the AGN feedback \citep{sorini_2021}. 
The median temperatures however do not change significantly; we see a small rise in typical temperature of \HI\ absorbers and a decrease for \OVI\ absorbers.  This occurs because the stronger photo-ionising background at $z=1$ results in slightly higher temperatures in the diffuse IGM, while less hot IGM gas owing to AGN feedback at higher redshifts results in lower \OVI\ temperatures.  

Earlier in this work we noted that the metal absorbers arise from a narrow range of gas temperature, which we now see does not significantly depend on redshift. In section \ref{sec:histograms} we showed that the absorber temperature is most sensitive to central galaxy stellar mass rather than star formation activity; the stellar mass of our galaxy sample does not change with redshift since we have selected our galaxies within the same stellar mass ranges at each redshift.
The typical metallicity of the absorbers increases with decreasing redshift, owing to a general build-up of metals over cosmic time due to star formation \citep{finlator_2008}. The largest increase in metallicity is seen in the \HI\ absorbers; where metal absorbers are seen they must arise from metal-enriched gas by necessity regardless of redshift, hence the increase in metallicity over cosmic time of the metal absorbers is less substantial. Thus the increase in typical metallicity of the metal lines reflects a further build-up of metals in gas that has already been enriched by $z=1$, whereas the large increase in typical metallicity of the \HI\ absorbers represents new metals that are added to the CGM at large since $z=1$.

In summary, the CDDF normalisation decreases with redshift for all absorber species we examine, owing to the large increase in comoving path length rather than a reduction in number of absorbers within the volume. The increase in metallicity over cosmic time somewhat counteracts the decrease in absorber overdensity such that there is still significant metal absorber populations at $z=0$. We note that these results are sensitive to our choice of galaxy sample selection with each redshift (a constant stellar mass criteria and redshift-dependent sSFR thresholds). This means that the quenched galaxies in our $z=1$ sample will have quenched via a different combination of mechanisms to our $z=0$ sample; likewise galaxies in our highest stellar mass bin at $z=1$ are more rare at that epoch. A complementary approach would be to identify the progenitors of our $z=0$ galaxy sample and examine the history of their CGM absorbers over the course of their individual evolutions; the progenitors would likely have higher absorber counts in their CGM owing to their lower masses at earlier epochs. 

\section{Contribution from satellites}\label{sec:satellites}

\begin{figure*}
    \includegraphics[width=\textwidth]{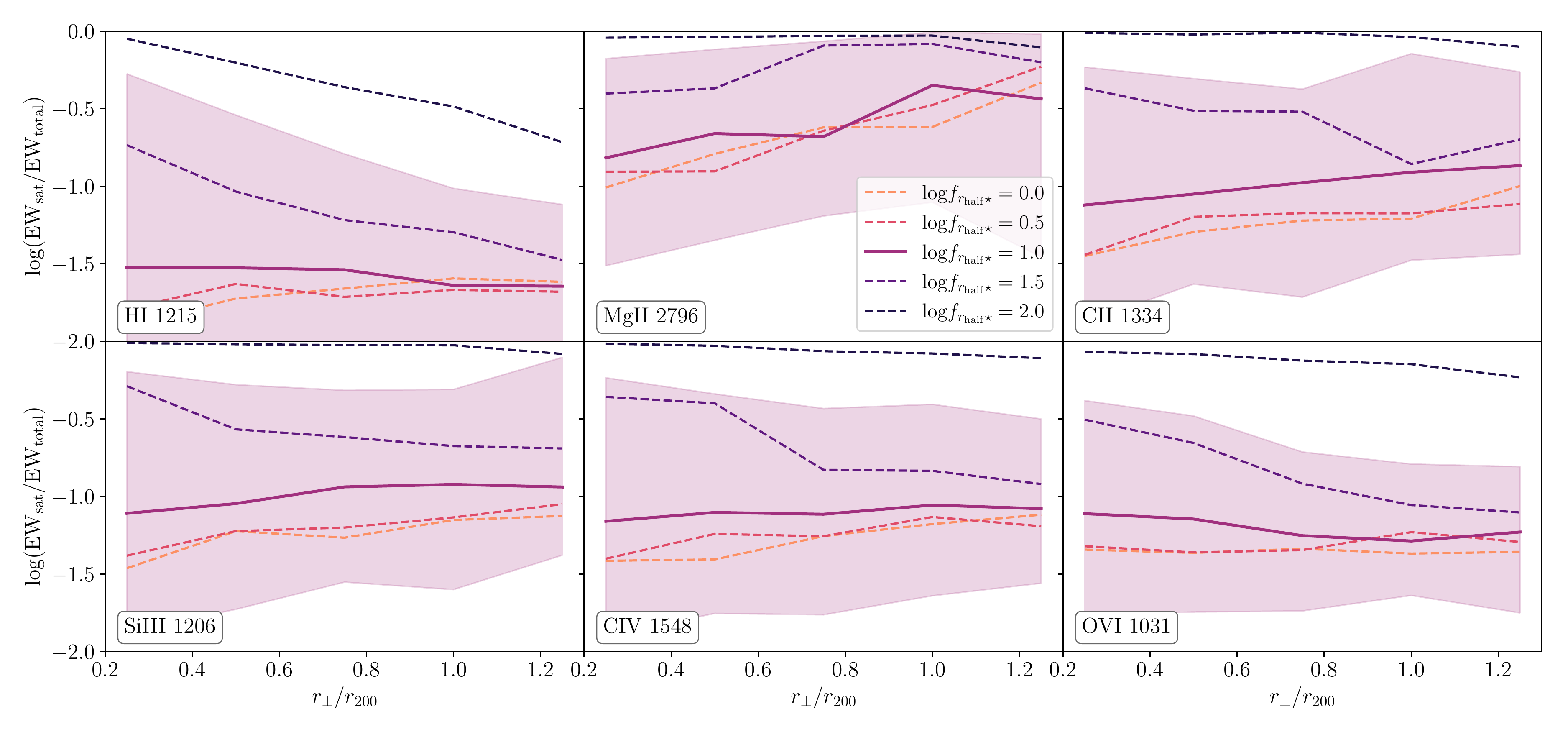}
	\vskip-0.1in
    \caption{Profiles of median fraction of equivalent width arising from particles associated with satellite galaxies. Lines are colour coded by the particle distance used to decide association with satellites, defined as factors of the satellite half mass radius ($f_{1/2,\star}$). Shading around the ${\rm log} f_{1/2,\star} = 1$ lines show the 25-75 percentile range.} 
    \label{fig:satellite_diffew}
\end{figure*}

A galaxy's halo contains not only the CGM but also satellite galaxies.  An interesting question is whether CGM's absorption at a given impact parameter is associated with a satellite galaxy rather than the halo gas of the central \citep[e.g.][]{agertz_2009, gauthier_2010, huang_2016}.  This is sensitively dependent on how one defines such an association.  In this section, we quantify the satellite contribution to CGM absorption, under various assumptions about how far out the gas from a satellite is considered to be associated.

We quantify the contribution to CGM absorption from satellite galaxies by regenerating the synthetic absorption spectra for our galaxy sample using only gas elements that are associated with satellites. We define associated satellite gas using the galaxy half stellar mass radius ($R_{{\rm half }\star}$).  That is, for each satellite, we identify all gas particles within various factors of $R_{{\rm half }\star}$, namely $\log\ R_{{\rm half }\star}=0,0.5,1,1.5, 2$. For each factor of $R_{{\rm half }\star}$ we output a new snapshot containing only particles associated with satellites within that distance, and generate new snapshot using the same lines of sight as in the main sample. A more theoretically-based approach to this problem would involve running a subhalo finder \citep[e.g. Rockstar, ][]{behroozi_2013}, and identifying all gas elements that are gravitationally bound to those sub halos; we leave this approach for future work, and focus here on our more observationally-accessible proxy.
The contribution of satellite galaxies to the original spectra is computed as the median fraction of equivalent width which arises from these new snapshots, with respect to the original spectral sample. We restrict this test to LOS that had at least one absorption region detected by our profile fitting procedure. 

Figure \ref{fig:satellite_diffew} shows profiles of median fraction of equivalent width arising from gas particles that are associated with satellite galaxies. Lines are colour-coded by the particle distance used to define association with satellites. The shaded region in each panel represents the 25-75 percentile range. The solid purple line in each panel represents the results for the satellite gas definition which we adopt going forward.

The satellite absorption fraction increases with increasing distance used to define satellite association, recovering essentially 100\% of the original equivalent width where particles are included up to $10^{2.5}\times$ and $10^{2}\times$ the satellite $R_{{\rm half }\star}$ for \HI\ and the metals, respectively. There is little increase in satellite contribution between snapshots using 1 and 3 times the satellite $R_{{\rm half }\star}$, and there is only a modest increase in satellite absorption fraction when increasing the radius from $1\to 10R_{{\rm half }\star}$, whereas above this the increase is substantial, suggesting that $10 R_{{\rm half }\star}$ is a reasonable and robust distance limit for gas particles to be associated with satellites. 
In Figure \ref{fig:satellite_diffew}, the results for this definition are shown as the solid line.
In this case, the satellite contribution to the overall \HI\ absorption is around 3\%, largely independent of radius.  Thus most of the \HI absorption is associated with ambient CGM gas, rather than satellites.

In contrast, for metals the contribution is always higher than this.  While satellite gas always provides a minority contribution to the overall absorption, it is evident that the low ionisation lines have a larger contribution from satellites than higher ionisation absorbers, and furthermore show a more strongly increasing trend with radius.
For instance, the satellite absorption fraction at 0.25$r_{200}$ increases from $\sim8\%$ for \OVI\ to $\sim16\%$ for \MgII, and the latter reaches $\sim40\%$ in the outer CGM.  This effect occurs since \MgII\ and \CII\ typically arise from cool, dense gas (see section \ref{sec:phase_space}) that is unlikely to be in the warmer ambient medium of the galaxy halo, unless that gas exists near a satellite galaxy. In contrast, high ionisation absorption like \OVI\ is spread more uniformly throughout the halo, leading to little radial trend and mostly coincident association with satellites.  In short, CGM absorption predominantly comes from ambient gas, but particularly for low ionisation absorption in the outer CGM, a substantial fraction is expected to arise from gas close to a satellite galaxy.

\section{Conclusions}

We have examined a population of low redshift CGM absorbers and their physical conditions in the \simba\ cosmological simulations. The underlying galaxy sample population was chosen to evenly sample a range of stellar masses and star formation activities at $z=0, 0.25, 0.5$ and $z=1$. For each galaxy in our sample we have generated realistic line of sight synthetic absorption spectra for \HI\ and a selection of metal ions at a range of impact parameters, and fitted Voigt line profiles to extract absorber column densities and physical conditions. We have used this sample to examine CGM absorber counting statistics in the form of the CDDF, the distribution of physical gas conditions that produces CGM absorption, and the redshift dependence of our results. In addition, we have estimated the contributions of collisional ionisation and of satellite galaxies to the resulting absorption strengths. Our main results are as follows:

\begin{itemize}
    
    \item The CGM CDDFs follow a power law at the high column density end, with a turnover at the low column density end which defines the completeness limit of our sample. \HI\ is the most abundant absorber species in our sample, with a large population with high column densities; \MgII\ is the least abundant absorber. The sample probes only a narrow range of \OVI\ column densities due to our spectrum generation approach.
    
    \item The normalisation of the CGM CDDFs (i.e. the number of absorbers at fixed redshift) increase with the sSFR of the central galaxies. The CDDFs from green valley galaxies resembles the quenched galaxy CDDF more closely than the star forming CDDF, suggesting that the CGM shows signs of quenching before the central galaxy itself is completely quenched. The CDDFs from high mass galaxies also have a lower normalisation owing to the hot gas content of these galaxy halos.
    
    \item \HI\ absorbers trace gas across a wide range of overdensities and temperatures owing to the high oscillator strength of \HI; metal absorbers trace a narrower range of physical conditions. Median gas overdensity decreases and temperature increases with increasing ion excitation energy. In general, median gas overdensity increases with increasing galaxy sSFR, while median gas temperature increases with increasing stellar mass.  The typical metallicity of CGM absorbing gas is around solar for all metal ions with only a weak radial trend, although \HI\ can trace significantly lower metallicity gas at larger $r_\perp$.
    
    \item Absorbers from the inner CGM are more abundant than those from the outer CGM, in general have higher column densities and trace denser gas.
    
    \item \HI\ absorption can arise from gas that is in any of the four regions of phase space (condensed, diffuse, hot gas, or WHIM). Condensed gas accounts for the majority ($\sim 100\%$) of all \HI and low metal ion absorption, and 87\% of \CIV absorption. \CIV, an intermediate ion, is the first of our selected metal ions to have a notable contribution from the other phase space regions (37\% of absorbers). In contrast with the other ions, \OVI absorbers predominantly arise from either hot gas or the WHIM (together 87\% of total absorption).

    \item The fraction of absorption from collisionally ionised gas is $\sim 25-55\%$ for \CIV\ and $\sim 80-95\%$ for \OVI. The collisional contribution increases with galaxy stellar mass and decreases with impact parameter.

    \item In general more overdense gas leads to higher column density absorption, up to ${\rm log} (N / {\rm cm}^{-2}) > 15$, above which the absorption features are saturated. 
    
    \item Absorbers around star forming galaxies trace higher overdensities and column densities, and have a more pronounced radial trend in overdensity than quenched galaxies, with higher overdensities and column densities in general occurring in the inner CGM. However the differences between star forming, green valley and quenched galaxies are small, and in general we find that absorbers around galaxies of different star formation activities do not probe fundamentally different phases of gas.
    
    \item When the excitation energy of the absorber species increases, the column density weighted median overdensity decreases, temperature increases, and metallicity increases to a lesser extent. Column density weighted overdensities and metallicities decrease with increasing impact parameter, however there is little trend with impact parameter and gas temperature.
    
    \item Although absorber counts increase with increasing redshift, this is counteracted by the increase in the comoving path length, resulting in a CDDF which decreases overall with increasing redshift. 
    
    %\item The column density weighted median overdensity decrease over cosmic time owing to stronger black hole driven feedback at low redshift. Column density weighted metallicities increase over cosmic time due to a build up of metals due to star formation and the enrichment of the CGM via star formation driven winds. The column density weighted temperatures of absorbers are largely unchanged with redshift owing to our choice of constant galaxy stellar mass with redshift for our underlying galaxy sample.
    
    %\item Regenerating the spectra without a photoionising background results in less overall absorption for each species. High excitation ions have the highest collisionally ionised fractions, with $\sim 30\%$ of \OVI\ absorption arising from collisional excitation for galaxies with $10 < {\rm log} (\mstar/\msolar) < 11$. Absorption that does occur arises from gas that is cooler, denser, and slightly less metal rich than absorbing gas in the photoionisation case.
    
    \item Satellite galaxies in the halos of central galaxies contribute $\sim 3\%$ of the overall \HI\ absorption and $\sim 10\%$ of metal absorption, except for \MgII\ absorption, of which $\sim 30\%$ comes from satellites. The fraction of \MgII\ absorption from satellites increases with increasing impact parameter. Overall, the fraction of metal absorption from satellites decreases with increasing ion excitation energy.  These results are based on a definition of satellite-associated gas being within 10 stellar half-light radii of the satellite, but are relatively insensitive to values below this.  This suggests that the dominant component of CGM absorption arises from ambient halo gas.

\end{itemize}

These results quantify how UV absorbers trace the typical physical conditions in the CGM around galaxies of different types.  It provides a guide towards interpreting low-redshift CGM absorption line observations, within the context of the \simba model.  In future work we plan to compare this to the results from other simulations, to understand how the different feedback processes implemented across present-day galaxy formation simulations impact CGM absorption tracers.

Our results show that the cool CGM gas is well-traced by low-ionisation metal absorbers, although all such species do not arise in the same gas.  This suggests that the common approach to obtaining physical conditions of CGM gas of fitting single-phase {\sc Cloudy} models to low-ionisation absorbers may not be fully accurate.  Also, the absorbing gas can be a highly biased tracer of the overall physical properties of CGM gas, so extracting information such as the total CGM mass (even within the cool phase) can be difficult.  The UV absorbers also clearly do not trace the dominant hotter gas component around quenched and green valley galaxies.  In the future we plan to investigate more comprehensive methods for extracting the physical conditions of the CGM.

Understanding the CGM and the role it plays in baryon cycle is a frontier issue in current galaxy formation models.  The work presented here is a step towards characterising the CGM via one particular technique, namely UV absorption.  Combining this with other approaches such as resolved \HI\ studies, X-ray absorption, and metal emission line mapping in the UV and X-rays will be critical for fully disentangling the complex multi-phase gas within the CGM of galaxies both today and back in time.

\section*{Acknowledgements}

We thank the anonymous referee for their valuable suggestions, which have improved the quality and clarity of this paper.
SA acknowledges helpful discussions with Massissilia Hamadouche, Rohit Kondapally and Britton Smith.
We thank Philip Hopkins for making \gizmo\ public, Robert Thompson for developing \caesar\ and Horst Foidl, Thorsten Naab and Bernhard Roettgers for developing \pygad.
SA is supported by a Science \& Technology Facilities Council (STFC) studentship through the Scottish Data-Intensive Science Triangle (ScotDIST).
RD acknowledges support from the Wolfson Research Merit Award program of the U.K. Royal Society.
Throughout this work, DS was supported by the European Research Council, under grant no. 670193, the STFC consolidated grant no. RA5496, and by the Swiss National Science Foundation (SNSF) Professorship grant no. 202671.
WC is supported by the STFC AGP Grant ST/V000594/1 and the Atracci\'{o}n de Talento Contract no. 2020-T1/TIC-19882 granted by the Comunidad de Madrid in Spain. He further acknowledges the science research grants from the China Manned Space Project with NO. CMS-CSST-2021-A01 and CMS-CSST-2021-B01.

\simba\ was run on the DiRAC@Durham facility managed by the Institute for Computational Cosmology on behalf of the STFC DiRAC HPC Facility. The equipment was funded by BEIS (Department for Business, Energy \& Industrial Strategy) capital funding via STFC capital grants ST/P002293/1, ST/R002371/1 and ST/S002502/1, Durham University and STFC operations grant ST/R000832/1. DiRAC is part of the National e-Infrastructure. 

For the purpose of open access, the author has applied a Creative Commons Attribution (CC BY) licence to any Author Accepted Manuscript version arising from this submission.

%%%%%%%%%%%%%%%%%%%%%%%%%%%%%%%%%%%%%%%%%%%%%%%%%%
\section*{Data and Software Availability}

The simulation data and galaxy catalogs underlying this article are publicly available\footnote{\url{https://simba.roe.ac.uk}}. The software used in this work is available on Github\footnote{\url{https://github.com/sarahappleby/cgm}} and the derived data will be shared on request to the corresponding author.



\bibliographystyle{mnras}
\bibliography{main} % if your bibtex file is called example.bib

%%%%%%%%%%%%%%%%%%%%%%%%%%%%%%%%%%%%%%%%%%%%%%%%%%

%%%%%%%%%%%%%%%%% APPENDICES %%%%%%%%%%%%%%%%%%%%%
\appendix

\section{Voigt fitting accuracy}\label{sec:appendix}

Here we provide a comparison of the results from our Voigt profile fitting routine to those retrieved directly from the synthetic spectra, both in terms of column density and equivalent width.

\begin{figure*}
    \includegraphics[width=\textwidth]{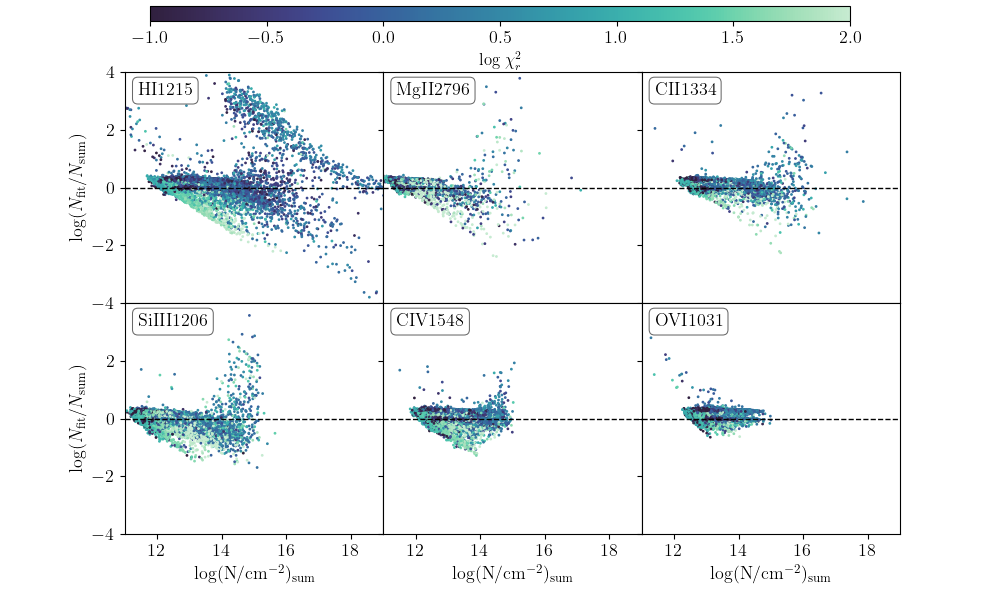}
	\vskip-0.1in
    \caption{Log difference between the column densities derived from spectrum fitting $N_{\rm fit}$ and column densities summed along the LOS $N_{\rm sum}$, against $N_{\rm sum}$. The column densities are totals within each absorption region, and the points are coloured by the $\chi_r^2$ of each region.} 
    \label{fig:compare_deltaN}
\end{figure*}

\begin{figure*}
    \includegraphics[width=\textwidth]{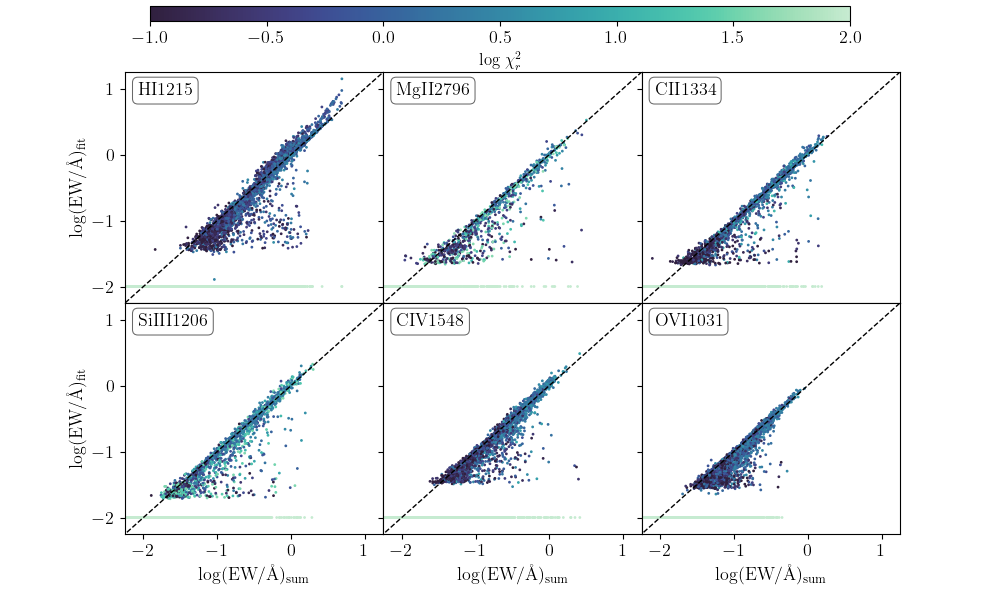}
	\vskip-0.1in
    \caption{The total fitted equivalent width along the LOS against the total equivalent width summed within the $\pm 600 \kms$ window along the LOS. Points are coloured by the maximum $\chi_r^2$ of the fitted regions along the LOS.} 
    \label{fig:compare_ew}
\end{figure*}

Figure \ref{fig:compare_deltaN} shows the log difference between the column densities derived from spectrum fitting $N_{\rm fit}$ and column densities summed along the LOS $N_{\rm sum}$, against $N_{\rm sum}$. The column densities presented here are totals within each detected absorption region. Points are coloured by the $\chi_r^2$ of each region; this dataset includes poor fits and column densities below the completeness limit. The horizontal dashed line at indicates where the fits match the summed column densities perfectly; the two measures of column density agree well in general. However, at the low (high) log$N$ end the fitting process has a tendency to underestimate (overestimate) the column density. In particular, the Voigt profile fitter gives an unreliable measure of \HI column density for Lyman Limit Systems since these features are saturated and lie in the logarithmic part of the curve of growth.

Figure \ref{fig:compare_ew} shows the total fitted equivalent width along the LOS (EW$_{\rm fit}$) against the equivalent width summed within the whole $\pm 600 \kms$ window (EW$_{\rm sum}$). In contrast to Figure \ref{fig:compare_deltaN}, here EW$_{\rm fit}$ and EW$_{\rm sum}$ are totals from the whole velocity window, rather than results for individual absorption regions. Here we include all LOS, including those with no detected absorbers (plotted at ${\rm log (EW/}$\AA$_{\rm fit}) = -2$). Points are coloured by the maximum $\chi_r^2$ of the fitted regions along the LOS. In general, EW$_{\rm fit}$ and EW$_{\rm sum}$ follow a 1:1 relation (indicated by the dashed black line). As with the column densities, Voigt fitting can overestimate the equivalent width of \HI\ at the high EW end. More commonly, the total equivalent width is underestimated due to one or more absorption regions not being detected and passed to the fitting routine.

%%%%%%%%%%%%%%%%%%%%%%%%%%%%%%%%%%%%%%%%%%%%%%%%%%

% Don't change these lines
\bsp	% typesetting comment
\label{lastpage}
\end{document}